\documentclass[usenatbib]{mn2e}
\usepackage{times}
\usepackage{graphicx} 
\usepackage{amsmath}
\usepackage{mathabx}
\usepackage{subfigure}
\usepackage{natbib}
\usepackage{hyperref}

\begin{document}

\title[Galaxy Structure and Quenching Efficiency]{The connection between galaxy structure and quenching efficiency}
\author[C. Omand, M. Balogh, B. Poggianti]
{Conor M. B. Omand$^1$$^,$$^2$, Michael L. Balogh$^{1,3}$, Bianca M. Poggianti$^4$  \\
$^1$Department of Physics and Astronomy, University of Waterloo, Waterloo, Ontario, N2L 3G1, Canada \\
$^2$Department of Physics and Astronomy, University of British Columbia, Vancouver, BC, V6T 1Z4, Canada \\
$^3$Leiden Observatory, Leiden University, PO Box 9513, 2300 RA Leiden, The Netherlands\\
$^4$INAF-Astronomical Observatory of Padova, l-35122 Padova, Italy}
\date{\today}
\maketitle
\begin{abstract}
Using data from the SDSS-DR7, including structural measurements from 2D surface brightness fits with GIM2D, we show how the fraction of quiescent galaxies depends on galaxy stellar mass $M_*$, effective radius $R_e$, fraction of $r-$band light in the bulge, $B/T$, and their status as a central or satellite galaxy at $0.01<z<0.2$.  For central galaxies we confirm that the quiescent fraction depends not only on stellar mass, but also on $R_e$.  The dependence is particularly strong as a function of $M_*/R_e^\alpha$, with $\alpha\sim 1.5$.  This appears to be driven by a simple dependence on $B/T$ over the mass range $9<\log(M_*/M_\Sun)<11.5$, and is qualitatively similar even if galaxies with $B/T>0.3$ are excluded.  
For satellite galaxies, the quiescent fraction is always larger than that of central galaxies, for any combination of $M_*$, $R_e$ and $B/T$.  The quenching efficiency is not constant, but reaches a maximum of $\sim 0.7$ for galaxies with $9<\log(M_*/M_\Sun)<9.5$ and $R_e<1$ kpc.  This is the same region of parameter space in which the satellite fraction itself reaches its maximum value, suggesting that the transformation from an active central galaxy to a quiescent satellite is associated with a reduction in $R_e$ due to an increase in dominance of a bulge component.
\end{abstract} 
\begin{keywords}
Galaxies: evolution, Galaxies: structure
\end{keywords} 

\section{Introduction}

It is now well known that galaxies in the local Universe can generally be separated into two broad classifications: blue, star-forming, disk galaxies and red, elliptical and lenticular galaxies with little or no recent star formation \citep[e.g.][]{Strateva,Kauffmann3,Baldry,Balogh2}.  The star formation rate (SFR) of blue galaxies is well correlated with their stellar mass (M$_*$), forming the ``main sequence'' or ``star-forming sequence'' in the SFR-M$_*$ plane \citep{Brinchmann,Salim,Gilbank,MBW+}.    Surveys show that the star-forming sequence has persisted out to z $=$ 1 \citep{Bell,Willmer,Noeske} and possibly z $=$ 2 and further \citep{Brammer,Taylor2}.  Most tellingly, the number density of red, passive galaxies has roughly doubled since z $=$ 1, while the number of blue galaxies has remained nearly constant \citep{Bell,Faber,Brown,Goncalves,Muzzin}.  This motivates the search for processes that cause blue, star-forming galaxies to transition to red, non-star-forming ones. This transition has become known as ``quenching''. 

It has also been established that the galaxy population varies with environment, in the sense that low-density environments are dominated by less massive, blue star-forming galaxies while denser groups and clusters tend to host massive red elliptical and lenticular galaxies \citep[e.g.][]{Oemler,DavisGeller,Dressler,Baldry2,Pasquali+10}.  
Interestingly, the star-forming sequence itself appears to be largely independent of environment, while the fraction of star-forming galaxies decreases dramatically with increasing local density \citep[e.g.][]{Carter,Balogh2,Wolf,P+08,Peng10}. 
\begin{table*}
\begin{tabular}{cccccccccccccc}
\hline
VAGC& Yang & Simard & RA & Dec & z&R$_e$&B/T&$M_*$&SFR& $m_r$&Central&C&V$_{\rm max}$\\
ID & ID & Rec. no. & \multicolumn{2}{c}{(J2000)} && (kpc)&  &($10^{10} M_\Sun$)& ($M_\Sun/$yr)&(mag)&flag&&($10^{6}$Mpc$^3$)\\
\hline
234667 & 1129345 & 432305 & 161.86430 & 13.889750 & 0.010 & 1.00 & 0.23  & 0.02 & 0.04 & 16.34 & 1 & 0.93 & 0.315\\
204290 & 958798 & 1095350 & 173.42712 & 50.449570 & 0.010 & 1.47 & 0.36 & 0.05 & 0.06 & 15.73 & 1 & 0.88 & 0.761 \\
114459 & 431764 & 897676 & 151.88120 & 57.729069 & 0.010 & 1.83 & 0.01 &  0.006 & 0.04 & 16.95 & 1 & 0.89 & 0.098 \\
256139 & 1219961 & 975660 & 178.49693 & 12.198710 & 0.010 & 1.87 & 0.17 &  0.08 & 0.15 & 15.08 & 1 & 0.86 & 1.875 \\
\hline
\end{tabular}
\caption{A sample of entries in our catalogue, available in the online version. These data are simply matched entries from the published catalogues of \citet{Blanton2005}, \citet{Brinchmann}, \citet{Simard}, and \citet{YangDR7}.  Columns 1-3 provide the ID for simple matching with \citet{Blanton2005}, \citet{YangDR7} and \citet{Simard}, respectively.  All catalogues were matched on RA, Dec and redshift (columns 4-6).  Column 7 is the circularized half-light radius from the \citet{Simard} bulge/disk decomposition model, with free index $n$ for the bulge component; column 8 is the corresponding fraction of light in the bulge.  Columns 9 and 10 are the stellar mass and SFR from \citet{Brinchmann}, and column 11 is the $r$-band petrosian magnitude from the original SDSS catalogue.  Column 12 is a flag that indicates whether (1) or not (2) a galaxy is the most luminous in its halo according to the \citet{YangDR7} catalogue; galaxies with a flag of 1 are considered a ``central'' in our analysis.  Finally, column 12 is the spatial completeness from \citet{Blanton2005}, and column 13 is the selection volume that we compute, based on the absolute magnitude and $R_e$ of the galaxy.  Additional columns that may be useful, including some that we make use of in the Appendix, are provided in the online table.\label{tab-catalogue} }.
\end{table*} 

There is some evidence to show that galaxy properties depend on the mass of their host, dark matter halo, with distinctly different behaviour for the most massive galaxies in the halo (known as centrals, since they are generally expected to lie near the centre of the halo) and the smaller, satellite galaxies that orbit around them  \citep{Hogg,Berlind,Pasquali+10}.  Some \citep{vdB,Peng12,Woo} suggest that the correlations with environment are due to transformations that take place when a galaxy first becomes a satellite.  This is natural, as such galaxies are expected to experience diminished cosmological gas accretion \citep{Balogh1,Kawata}, ram-pressure with the intrahalo gas \citep{Gunn,Grebel}, and a different type of interaction with neighbouring galaxies \citep{Moore1,Moore2} and the tidal field \citep{Read}.

The fraction of quiescent galaxies clearly correlates with galaxy stellar mass and this, combined with the apparent ease with which stellar masses can be computed from photometric data, has motivated modelers and observers to consider stellar mass as the fundamental parameter that determines a galaxy's properties \citep[e.g.][]{Peng10,Woo}.  However, 
various authors \citep[e.g.][]{Brinchmann,Kauffmann2,Franx} found that galaxy surface density is more predictive of the specific SFR (sSFR) than stellar mass.  Others have shown the correlation is best with velocity dispersion, either as measured directly from dynamics \citep[e.g.][]{SLH,Wake2,Wake1}, or inferred from the virial relations between mass and size 
\citep{Franx}. 
There are good indications that the trends are driven by conditions in the centre of galaxies, either as inferred from  S\'{e}rsic indices derived from single-profile fits to the surface brightness \citep{Blanton2003c,Driver,Allen,Schiminovich,Bell2,Donofrio+11}, central surface density \citep[e.g.][]{Cheung,Fang}, or presence of a central bulge \citep{Martig,Cappellari2}.
While it remains controversial which of these characterisations is most fundamental, it seems clear that galaxy structure, and not just stellar mass, is relevant \citep[e.g.][]{P+13}.  

If galaxy properties are primarily determined by some aspect of their structure, rather than their stellar mass alone, it reopens the question of how (or if) galaxies of a given {\it structure} depend on their environment.  The first purpose of this paper is to quantify the quenching efficiency of satellite galaxies as a function of their structure.  This leads us to take a more general approach than some of the studies listed above, and consider how the quiescent fraction of central galaxies depends on galaxy stellar mass and radius in the local Universe sampled by the Sloan Digital Sky Survey \citep[SDSS,][]{SDSS}.  

All calculations use the $\Lambda$CDM cosmology consistent with the nine year data release from the WMAP mission, with parameters $h$ = 0.693, $\Omega_M=0.286$, and $\Omega_\Lambda=0.714$ \citep{WMAP}.
\section{The Data}
\subsection{The Sample}
We draw our sample from the seventh data release of the SDSS.  The redshift range is limited to $0.01 < z < 0.2$.  We extract data from catalogs made by \citet{Blanton2005}, \citet{Brinchmann}, \citet{Simard}, and \citet{YangDR7}.  All catalogs were position matched to the closest galaxy within an arcsecond and redshift matched to within 0.0005.  The resulting sample size after matching is 581 990, but this is reduced to 471 554 after making the completeness cuts described in Section~\ref{scn:cvc}.  This matched catalogue, with all quantities relevant to reproduce the results in this paper, is made available online; with sample entries shown in Table~\ref{tab-catalogue}.
We use the DR7 group catalog from \citet{YangDR7}, which was generated by grouping galaxies with SDSS redshifts using the same friends-of-friends algorithm as their original DR4 catalog \citep{Yang1}.  
We define the most luminous galaxy in each group to be the central galaxy and other members to be satellites; however the changes to our results are insignificant if we define the most massive galaxy to be the central.  Isolated galaxies (those not linked to any group) are also defined as centrals.  Using this definition, we obtain 376 028 centrals and 95 526 satellites following the cuts described in Section~\ref{scn:cvc}.

\subsection{Galaxy Properties} \label{scn:galprop}
We adopt the Petrosian $r$ magnitudes, $m_r$ for all galaxies \citep{DR7}, and calculate the absolute magnitudes from these as 
\begin{equation}
M_r = m_r - DM(z) - k_{0.1}(z)+1.62(z-0.1),
\label{eqn:abmag}
\end{equation}
where $DM(z)$ is the distance modulus, 
$1.62(z-0.1)$ is the evolution term from \citet{Blanton2003b}, and 
$k_{0.1}(z)$ is an approximation
of the k-correction to z=0.1 \citep{Yang3,Blanton2003a,BlantonRoweis}, given by
\begin{equation}
k_{0.1}(z) = 2.5log\left({\frac{z+0.9}{1.1}}\right).
\label{eqn:kcorr}
\end{equation}
We convert these absolute magnitudes to luminosities using M$_\Sun=$ 4.64.

Measurements of stellar mass ($M_*$) and SFR are taken from the online catalogue provided by Brinchmann et al. (http://www.mpa-garching.mpg.de/SDSS/DR7/).  These measurements are based on the procedure described by \citet{Brinchmann}, and assume a \citet{Kroupa} initial mass function.  The SFR within the fiber is determined directly from the H$\alpha$ and H$\beta$ line emission in star forming galaxies, and from the D4000 break for galaxies with weak emission or evidence for AGN.  Outside the fiber, using a method similar to \citet{Salim}, models of stellar populations are fit to the observed photometry.  The SFRs from the fiber were added to the SFRs from the photometry fits to give an estimate of the total.  The uncertainty of these SFRs is estimated to be around 0.2 dex for star forming galaxies, and 0.7 dex or more (in fact, best treated as upper limits) for quiescent galaxies \citep{Woo}.
Stellar masses are measured with a spectral energy distribution fitting method similar to the one used by \citet{Salim} to measure SFR.

The effective radii R$_e$ we use are the 
r-band circular half-light radii
from  \citet{Simard}.  The parameters are derived using the GIM2D software \citep{GIM2D} to fit the surface brightness profiles with parametric models.  These fits were performed simultaneously on $g$ and $r$ images, using a simple disk or bulge $+$ disk model, with particular care given to subtraction of the background, which can be a large source of systematic uncertainty.
Three model fits are provided: an n$_b=$ 4 (de Vaucouleurs) bulge $+$ disk decomposition, a free n$_b$ bulge $+$ disk decomposition, and a single-component S\'{e}rsic model.  We take our parameters from the free n$_b$ bulge $+$ disk decomposition, except where otherwise noted\footnote{Specifically, when we consider properties of the bulge component we use the models with $n_b = 4$.}, and we justify this decision in Appendix~\ref{scn:ressel}.

In this paper, we will consider correlations with galaxy stellar mass, ``inferred velocity dispersion''\footnote{The stellar velocity dispersion is assumed to be related to $M_*$ and $R_e$ through the virial relation $\sigma^2$ $ \propto$ $GM_*/R_e$} M$_*$/R$_e$, and surface density\footnote{Surface density is actually M$_*$/$\pi R^2_e$, but we omit $\pi$ for simplicity.} M$_*$/R$_e^2$.  In Appendix~\ref{scn:IVDSDF} we present several figures which illustrate how the sample is distributed amongst these parameters.  While not directly related to the results in the rest of the paper, they can be helpful for interpreting some of the trends we discuss.
\subsection{Completeness Corrections} \label{scn:cvc}
\subsubsection{Magnitude Limit}

The SDSS Legacy survey targets objects with $m_r < 17.77$, which means the stellar mass limit of the sample is redshift dependent.  First, we calculate a conservative absolute magnitude limit,
\begin{equation}
M_{r,lim}=17.77-DM(z)-k_{0.1}(z)+1.62(z-0.1)-0.1,
\label{eqn:mglim}
\end{equation}
where the extra $-0.1$ factor is brought in to account for the scatter in k-correction \citep{Yang3}. The few galaxies with $M_r$ fainter than this limit are removed from the sample.  
For the remaining galaxies, we calculate the maximum redshift for which the galaxy would be brighter than $m_r=17.77$, 
and calculate the corresponding selection volume, $V_{\rm max}$, as
\begin{equation}
V_{max}=0.1947\times\frac{4}{3}\pi\frac{d_C(z)^3 - d_C(z=0.01)^3}{(1+z)^3},
\label{eqn:Vcom}
\end{equation}
where the factor 0.1947 represents the fraction of sky covered by the SDSS \citep{Simard}, and $d_C$ is the comoving distance.  
\begin{figure}
  \begin{center}
  \includegraphics[clip=true,trim=0mm 0mm 0mm 0mm,scale=0.55,angle=0]{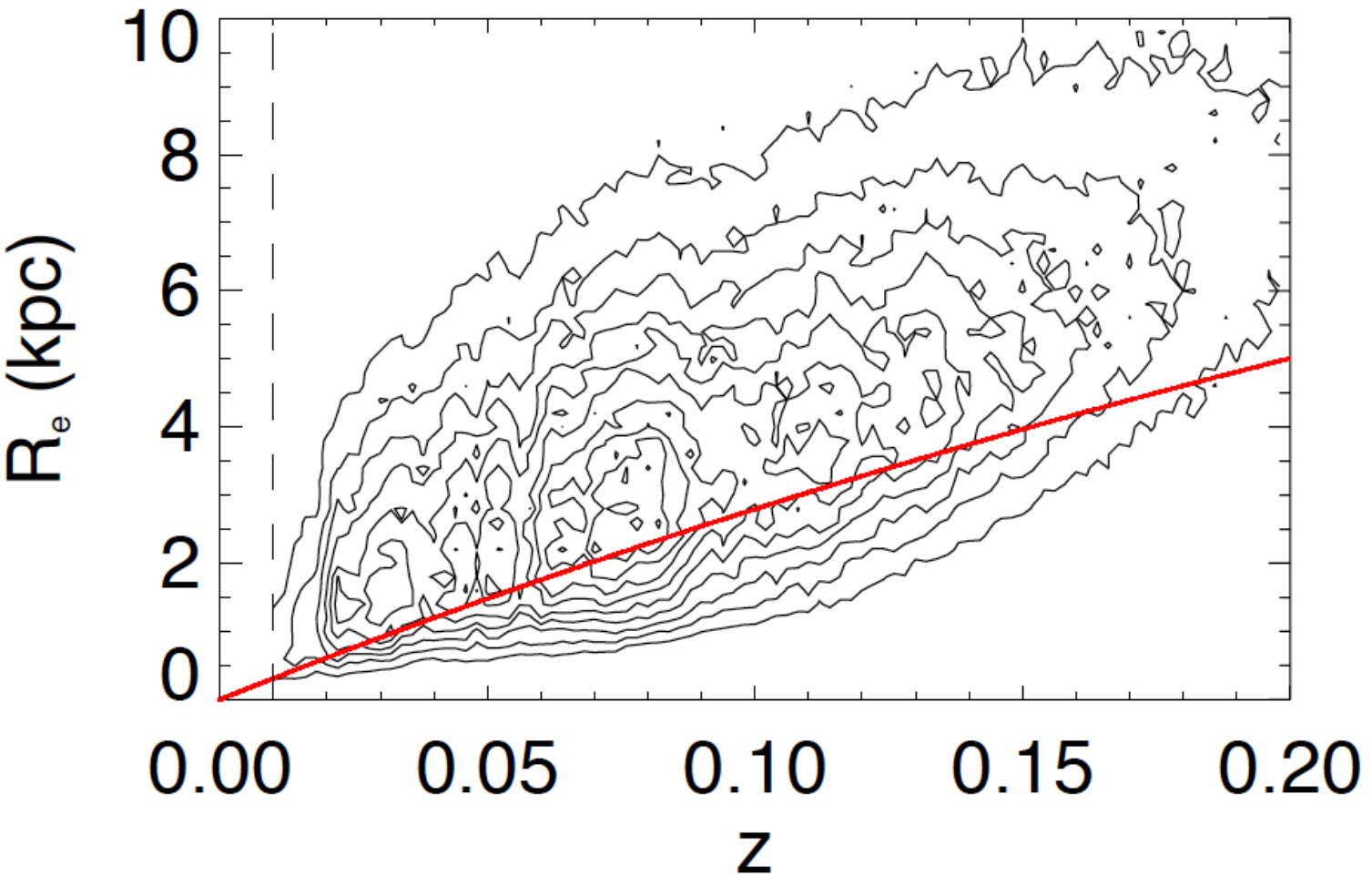}
  \caption{The effective radii of galaxies in our initial SDSS sample are shown as a function of their redshift.  Instead of showing individual points, we use logarithmically-spaced contours to represent the density of points in this space.  The red line indicates the radius corresponding to an angular size of 1.5$\arcsec$ as a function of redshift.  Galaxies below this limit are removed from the sample to define a clean selection limit.  } 
  \label{fig:relim}
 \end{center}
\end{figure} 

\subsubsection{Angular Resolution} \label{scn:AR}
Figure~\ref{fig:relim} shows the distribution of $R_e$ as a function of redshift for all galaxies in our initial sample.  The apparent correlation is driven primarily by the fact that the limited spatial resolution of the SDSS means that physically small galaxies will appear as point sources and will not be selected for spectroscopy in the main galaxy sample \citep{Strauss}.  
The average angular resolution of SDSS is $\sim$ 1.2$\arcsec$ \citep{Shin}, but we take a conservative minimum of 1.5$\arcsec$ for this calculation. The actual galaxy selection criteria are more complicated than this \citep{Scranton,Taylor}, but our conservative approach admits a simpler analysis.  The red line in Figure~\ref{fig:relim} shows the value of $R_e$ corresponding to an angular size of 1.5$\arcsec$, as a function of redshift.  To make a well-defined sample, we exclude galaxies below this line from the rest of the analysis.  The selection volume of the remaining galaxies may be limited by the distance at which they would drop below this line, rather than by their luminosity.  Specifically, for each galaxy we calculate the redshift at which the angular size of its effective radius is 1.5",
and calculate the corresponding volume correction from Equation~\ref{eqn:Vcom}.

\subsubsection{Weights}
The selection volume of each galaxy is taken to be the smaller of the two volumes described above.  In addition, we weight galaxies by the spatially-dependent spectroscopic completeness factor, $C$,  available from the NYU-VAGC \citep{Blanton2005}.  Thus, each galaxy in our sample is given a weight of
\begin{equation}
w_i=\frac{1}{V_{max}C},
\label{eqn:Vmaxweight}
\end{equation}
which represents the number of such galaxies per unit volume in a complete sample.

\subsection{Separating Active and Quiescent Galaxies}
\begin{figure}
\centerline{\includegraphics[clip=true,trim=0mm 0mm 0mm 0mm,scale=1.,angle=0]{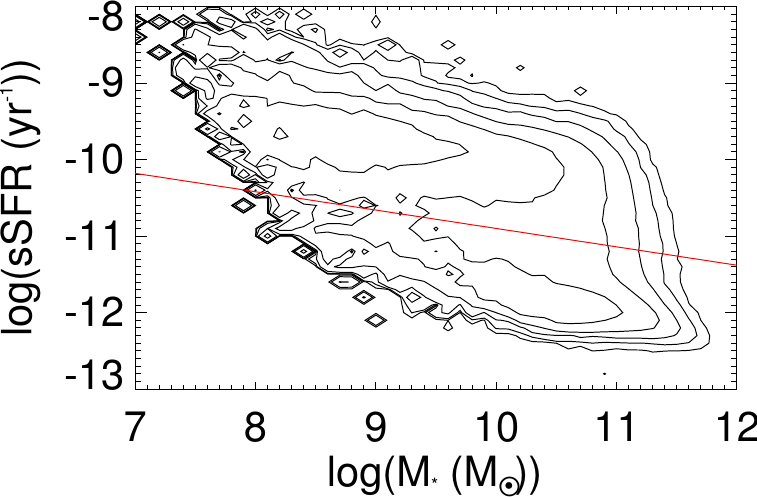}}
        \caption{The sSFR is shown as a function of stellar mass, with contours representing the weighted number density of points in this plane.  The red line represents our division into quiescent and active galaxies.}
\label{fig:idf}
\end{figure}

We base our separation of star-forming (active) and non-star-forming (quiescent) galaxies on the sSFR of each galaxy.  Figure~\ref{fig:idf} shows the correlation of sSFR and stellar mass, where the contours indicate the weighted density of the population.
The main sequence of star-forming galaxies is readily apparent, and clearly separated from the rest of the galaxy population (recall that SFR is not accurately measured at low values, where it is best treated as an upper limit).  This bimodality motivates dividing the population into two and, since the main sequence shows a small mass-dependence, our division must also be mass dependent.  We divide the population at $sSFR_o(M_*)$, where
\begin{equation}
\log{(sSFR_o)} = -0.24\log{(M_*/M_\Sun)} - 8.50.
\label{eqn:cSMdiv}
\end{equation}
Any galaxy with sSFR $<$ sSFR$_o$ is considered quiescent.  This division is shown on Figure~\ref{fig:idf} as the red line. 

\subsection{Stellar Mass and Effective Radius Distribution}  \label{scn:dd}
\begin{figure}
\includegraphics[scale=1,angle=0]{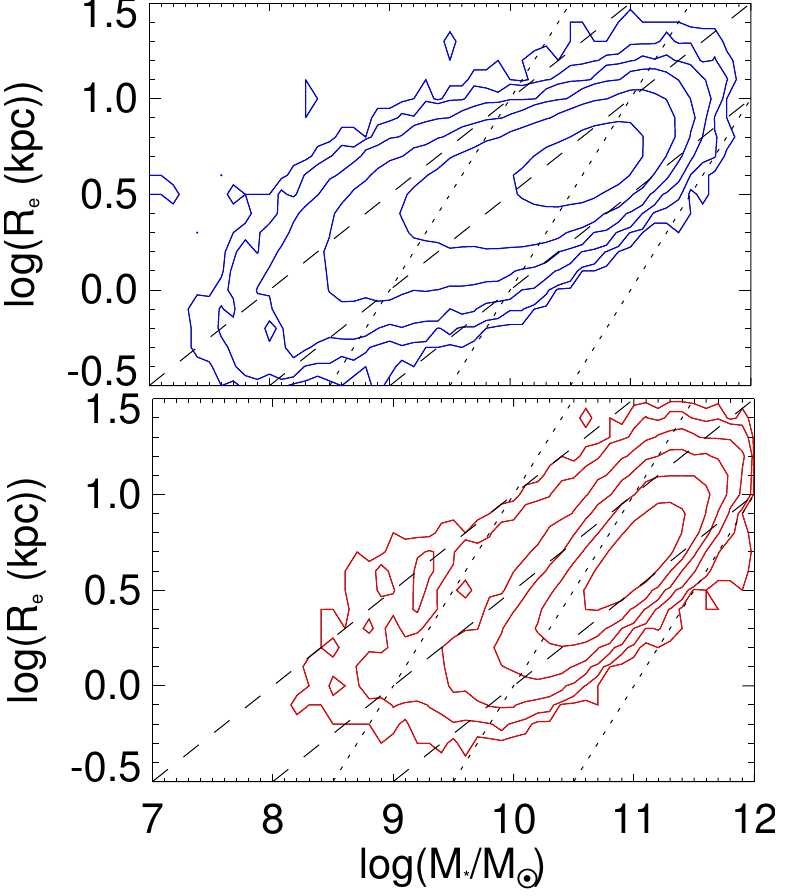}
	\caption{Contours show the raw (unweighted) number density of central galaxies as a function of stellar mass M$_*$ and effective radii R$_e$, for active (top) and quiescent (bottom) galaxies.   Contours are logarithmically spaced with 1.5 contours per dex.  The dashed lines represent surface densities of 10$^8$, 10$^9$, and 10$^{10}$ M$_\Sun$/kpc$^2$ from left to right, while the dotted lines represent inferred velocity dispersions of 10$^9$, 10$^{10}$, and 10$^{11}$ M$_\Sun$/kpc, for reference. }
\label{fig:ud}
\end{figure}
We now present the distribution of the quiescent and active galaxy populations as a function of both their stellar mass and effective radius.  Figure~\ref{fig:ud} shows the raw distribution of active and quiescent central galaxies in this plane, where the contours indicate the number of galaxies in the sample.  There is a well-defined correlation between these quantities, with a slope that is clearly steeper for quiescent galaxies than it is for star--forming galaxies.  For reference, we show lines of constant surface mass density ($M_*/R_e^2$, dashed lines) and constant inferred velocity dispersion ($M_*/R_e$, dotted lines).   

\begin{figure}
		\includegraphics[scale=1,angle=0]{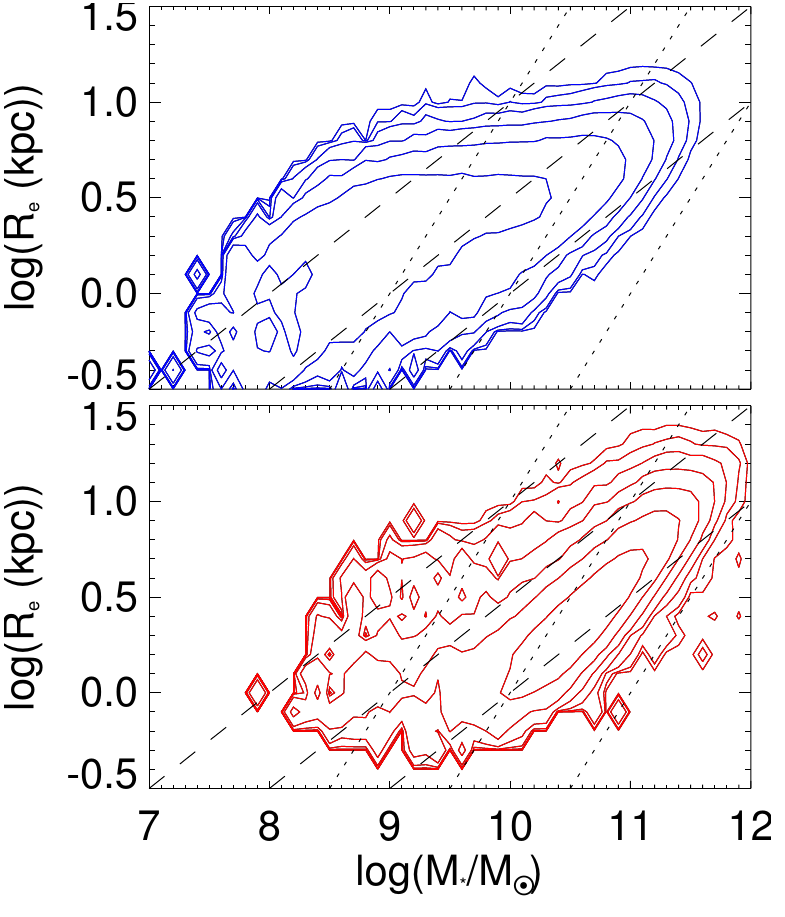}
	\caption{As Figure~\ref{fig:ud}, but where the contours represent the weighted density; equivalently, the physical volume density of central galaxies. Contours are again logarithmically spaced, with 1.5 contours per dex.}
\label{fig:wdc}
\end{figure}

Figure~\ref{fig:ud} is useful to see where the data actually lie.  However, because the galaxy selection is biased against low-luminosity and low-$R_e$ galaxies at higher redshift, the physical distribution is obtained by weighting the galaxies as given in Equation~\ref{eqn:Vmaxweight}: this weighted distribution is shown in 
 Figure~\ref{fig:wdc}.  This naturally has a large effect on the distribution.  It is still evident that the $M_*-R_e$ relation is fundamentally different for the active and quiescent populations, as is well-known \citep[e.g.][]{Kormendy85,P+13}.  Star--forming galaxies show a scaling which is approximately $M_*$ $\propto $ $R_e^2$ (dashed lines), representing constant surface mass density.  The massive, quiescent galaxies show a considerably steeper relation, that lies somewhere between lines of constant mass density and constant inferred velocity dispersion (the latter indicated by dotted lines).  At lower masses, $M_*<10^{10}M_\Sun$ or so, the volume density of central, quiescent galaxies drops off rapidly, in agreement with \citet{Geha}.  Moreover at these masses the correlation becomes flatter, with $R_e$ nearly independent of mass.

\begin{figure}
		\includegraphics[scale=1,angle=0]{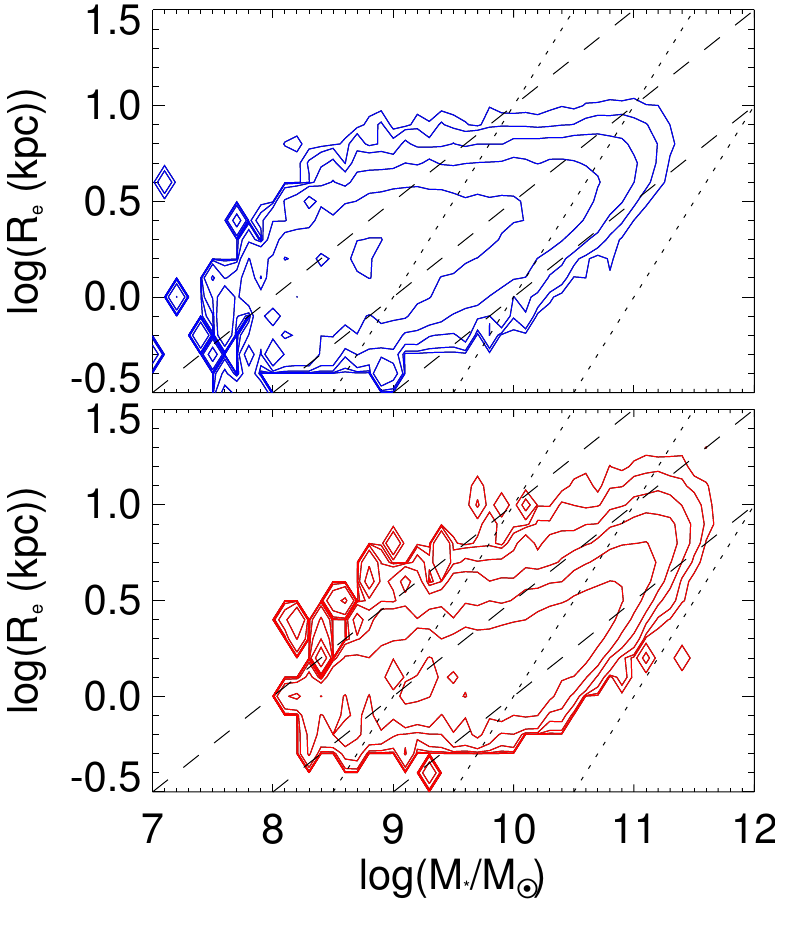}
	\caption{As Figure~\ref{fig:wdc}, but for satellite galaxies.}
\label{fig:wds}
\end{figure}

We show the same distributions for satellite galaxies, in Figure~\ref{fig:wds}.  Interestingly, the population of active satellite galaxies shows a very similar correlation between $M_*$ and $R_e$ to the one that is observed for central galaxies.  However, the distribution of quiescent galaxies is markedly different. In particular, the relative number of low-mass ($M_*<10^{9.5}M_\odot$), quiescent galaxies is larger for satellite galaxies, compared with centrals; this is most easily seen as a steeper stellar mass function for quiescent satellites (see Figure~\ref{fig:smf}).  These low-mass galaxies do not follow the tight correlation between $M_*$ and $R_e$ exhibited by more massive galaxies, but have approximately uniform sizes of $R_e\sim 1$ kpc.

The minimum observable effective radius in the data is R$_{\rm min}=0.3$ kpc, corresponding to $1.5$\arcsec\ at the lowest redshift of our sample, $z=0.01$.  Figure~\ref{fig:wdc} shows that the distribution of both populations approaches this limit. In particular, our sample of star-forming galaxies may be incomplete for $M_*<10^8M_\Sun$, where the size distribution indicates that a substantial fraction of galaxies likely lie below our resolution limit.  The quiescent galaxies appear to reach a minimum radius of $\sim 0.5$ kpc, for galaxies with masses $M_*<10^{10}M_\Sun$.  This is, perhaps, uncomfortably close to the selection limit of R$_{\rm min}=0.3$ kpc.
However, this distribution is in agreement with results shown by 
\citet{Kormendy}, using higher resolution data.  
They demonstrate (see their Fig. 38) that the lower surface brightness distribution of spheroidal galaxies indeed reaches a minimum of $\sim 0.4$ kpc.  Their data do indicate that the high surface-brightness, elliptical galaxy sequence may extend to smaller radii; extrapolating the relation we observe in Figure~\ref{fig:wdc} shows this limit would be reached for galaxies with $M_*<10^9M_\Sun$.  However, the mass function of these elliptical galaxies peaks well above this point.  
An {\it HST} study of the Coma Cluster by \citet{Hoyos} also demonstrates that there are few, if any, galaxies with $R_e<R_{\rm min}$.
Thus, formally we expect our sample is only fully complete for $M_*>10^9M_\Sun$, although significant incompleteness is unlikely to be an issue for up to an order of magnitude below that limit.  
\begin{figure}
\centerline{
		\includegraphics[scale=1,angle=0]{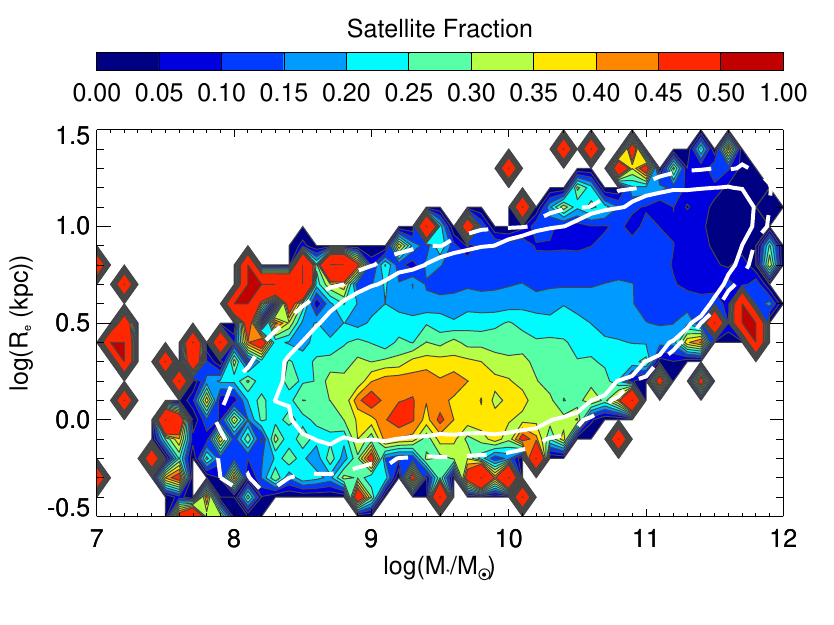}
}
	\caption{The satellite fraction as a function of 
stellar mass M$_*$ and effective radii R$_e$. Differences between coloured contour levels are statistically significant within the solid white contours, which indicate the region where the number density of galaxies contributing to the calculation is greater than 100 per bin.  The dashed, white contours indicate a lower number density of 20 per bin; correlations outside this region are of low statistical significance.}
\label{fig:fsat}
\end{figure}\begin{figure*}
\centerline{
                \includegraphics[clip=true,trim=0mm 0mm 0mm 0mm,scale=1.5,angle=0]{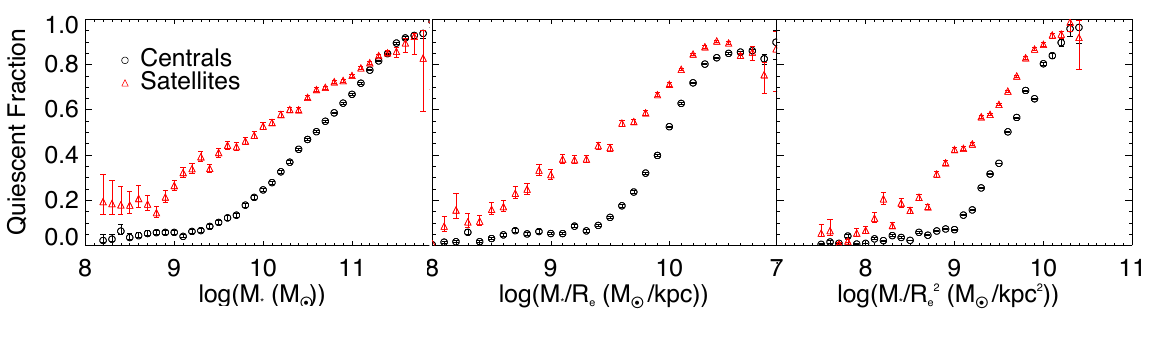}
}
        \caption{The quiescent fraction of galaxies in our sample is shown as a function of stellar mass (left), inferred velocity dispersion (middle), and surface density (right).  Central galaxies are shown as black circles, while satellites are represented by red triangles.  Error bars represent the $1\sigma$ statistical uncertainty.}
\label{fig:qfsis}
\end{figure*}
\begin{figure*}
\centerline{
               \includegraphics[clip=true,trim=0mm 0mm 0mm 0mm,scale=1.25,angle=0]{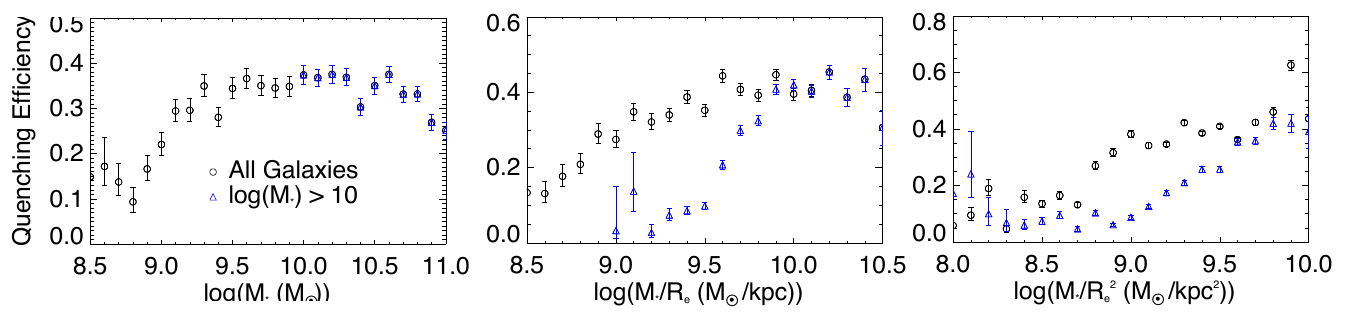}
}
        \caption{The quenching efficiencies (Equation~\ref{eqn:qe}) of galaxies as a function of stellar mass (left), inferred velocity dispersion (middle), and surface density (right).  Results for all galaxies are shown as black circles, and results for galaxies with M$_* > 10$ are shown as blue triangles.}
\label{fig:qe1d}
\end{figure*}

 Figure~\ref{fig:fsat} shows the fraction of all galaxies that are satellites, as a function of $M_*$ and $R_e$.  
Satellites show a different distribution in both mass and radius than central galaxies; they are generally smaller and less massive.  Their relative abundance peaks at $\sim 50$ per cent, around M$_*$ $=$ 10$^{9.2} M_\Sun$ and R$_e$ $=$ 1 kpc.  In this figure, and similar figures that follow, we indicate the region where the galaxy density exceeds 100 per bin by the white, solid contours.  Within this region, uncertainties on all fractions should be significantly less than the spacing between contour levels representing the quantity of interest.  The dashed line shows a lower density of 20 per bin; regions outside the dashed line are of low statistical significance.
We confirm that the mass-dependence of the satellite fraction is in good agreement with that given by \citet{Wetzel}, as expected since we are using the same sample and definition.  It is clear by comparing with Figures~\ref{fig:wdc} and \ref{fig:wds} that this maximum in satellite fraction is due to the population of high-surface brightness, quiescent galaxies that is largely absent among central galaxies.  

\section{Results}  \label{scn:res}

\subsection{Quenching and Quenching Efficiency}
We start by showing the fraction of central and satellite galaxies that are quiescent, as functions of stellar mass, inferred velocity dispersion, and surface density, in Figure~\ref{fig:qfsis}.  The 1$\sigma$ uncertainties displayed are calculated assuming the relative uncertainty of the weighted fraction is the same as that calculated on the unweighted distribution determined from the beta distribution of \citet{Cameron}. 
As previously shown by others \citep[e.g.][]{Kauffmann2,Cheung, Wake2,Wake1}, we find that the quiescent fraction 
increases less sharply with stellar mass than with either inferred velocity dispersion or surface density.  To quantify this, we note that the quenched fraction of central galaxies increases from $0.2$ to $0.8$ over 1.4 dex in $M_*$, compared with only $\sim 0.7$ dex in the other two quantities. 
 
The red points in Figure~\ref{fig:qfsis} show that the quiescent fraction of satellite galaxies is {\it always} larger than it is for centrals of the same stellar mass, surface density, or inferred velocity dispersion.   In general, the quiescent fraction of satellites has both a higher normalization and is a weaker function of these variables than it is for central galaxies.  Thus means the difference between the satellite and central populations tends to decrease as these measures increase.

\citet{Baldry2}, \citet{Peng10} and others have shown that the trend of increasing quiescent fraction with increasing local density is largely independent of stellar mass.  It is also fundamental in the hierarchical CDM model that satellite galaxies were once centrals in their own halo.  These motivate us, following \citet{vdB} and \citet{Peng10}, to consider the efficiency with which star formation is quenched in galaxies following their transformation from central to satellite.  Physically, this is the fraction of satellites that were active when they became satellites and were subsequently quenched by the satellite-exclusive process \citep{vdB}.  Operationally, if we assume that the progenitors of the satellite population are identical to the local population of central galaxies, and if the associated quenching process does not significantly change the stellar mass of the galaxy, then we can measure this efficiency from the quiescent fraction of satellite ($QF_{\rm sat}$) and central ($QF_{\rm cent}$) galaxies as
\begin{equation}
QE=\frac{QF_{sat}-QF_{cent}}{1-QF_{cent}}.
\label{eqn:qe}
\end{equation}
This quantity is presented in Figure~\ref{fig:qe1d}, as a function of all three parameters under consideration here.    In all three cases, this efficiency is approximately constant at $\sim 0.4$, at least above some threshold value.  This reflects the fact that high stellar mass (or high surface density, etc.) galaxies are more likely to be quiescent as a result of their ``intrinsic'' properties, regardless of whether or not they are satellites.  The relatively small difference in quenched fraction between satellites and centrals at these high values is thus still indicative of a highly efficient process.

We note that galaxy samples that are stellar mass limited, including our own, may not be complete in $M_*/R_e$ or $M_*/R_e^2$, as is apparent from Figures~\ref{fig:wdc} and \ref{fig:wds}.   A selection limit in $R_e$ further complicates this, and the level of completeness depends on how the mass-size relation extrapolates to arbitrarily low masses.   Our sample is limited to $M_*\gtrsim 10^{7.5} M_\odot$ and, assuming there is not a large population of galaxies with sizes below our resolution limit of $\sim 0.3$ kpc, Figure 4 shows that we are also likely complete for $M_*/R_e>10^{8}M_\odot/$kpc.  However, as the $M_*-R_e$ relation for star--forming galaxies is approximately parallel to lines of constant $M_*/R_e^2$, this quantity is potentially incomplete for all values. We show the effect of imposing a higher stellar mass limit by comparing, as blue triangles on ~\ref{fig:qe1d}, the results for a sample restricted to $M_*>10^{10}M_\Sun$.  This significantly changes the shape of the dependence on structural variables, with QE only remaining unchanged at the highest values of $M_*/R_e$.  This highlights the difficulty in interpreting 1D projections such as those shown in Figure~\ref{fig:qfsis}, and this partially motivates a more general approach.

\begin{figure*}
		\includegraphics[scale=1.8,angle=0]{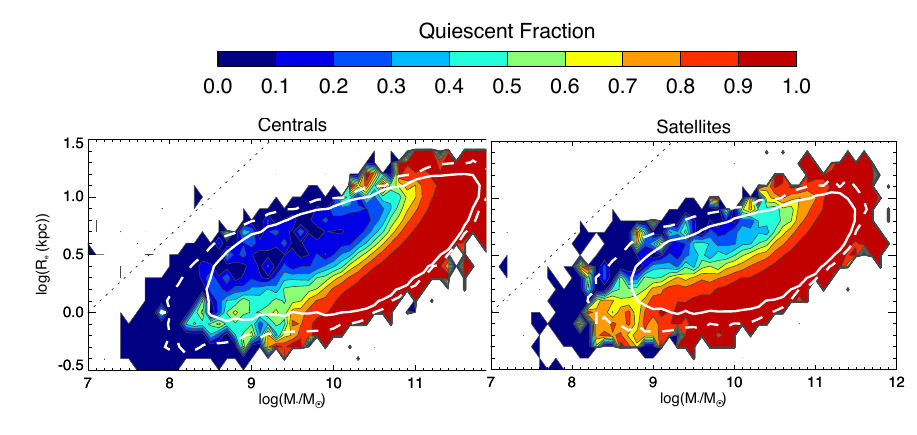}
	\caption{The central (left) and satellite (right) galaxy quiescent fraction as a function of stellar mass M$_*$ and effective radius R$_e$.  Contours are spaced every 0.1 in quiescent fraction.  The dashed line indicates constant M$_*$/R$_e^{1.5}$. Solid and dashed, white contours indicate the region where the number density of galaxies contributing to the calculation is greater than 100 per bin, and 20 per bin, respectively.  This representation demonstrates that the quiescent fraction does not depend only on stellar mass, but also on galaxy structure.}
\label{fig:csqf}
\end{figure*}

While physically motivated, the combinations of $M_*/R_e$ and $M_*/R_e^2$ are somewhat arbitrary, and we now consider the quiescent fraction in a more general way, as a function of $\log({M_*/M_\Sun})$ and $\log({R_e})$ independently.  Figure~\ref{fig:csqf} shows contours of constant quiescent fraction in this plane.
For central galaxies, we observe the remarkable fact that the contour lines are closely spaced and nearly linear in this space, following an approximate relation $M_*$ $ \propto $ $R_e^{1.5}$, shown by the dashed line.  It is quite evident from this representation that, at least for masses $M_*\lesssim10^{10.5}M_\Sun$, the quiescent fraction does not depend on stellar mass alone.  

\begin{figure}
\centerline{
		\includegraphics[scale=1,angle=0]{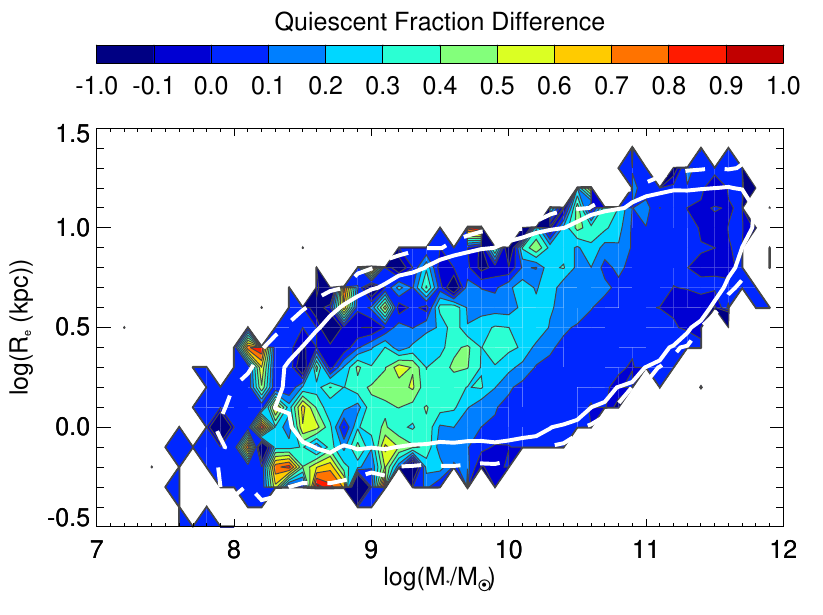}
}
	\caption{The quenched fraction difference  as a function of stellar mass M$_*$ and effective radius R$_e$. }
\label{fig:qfd}
\end{figure}
\begin{figure}
\centerline{
		\includegraphics[scale=1,angle=0]{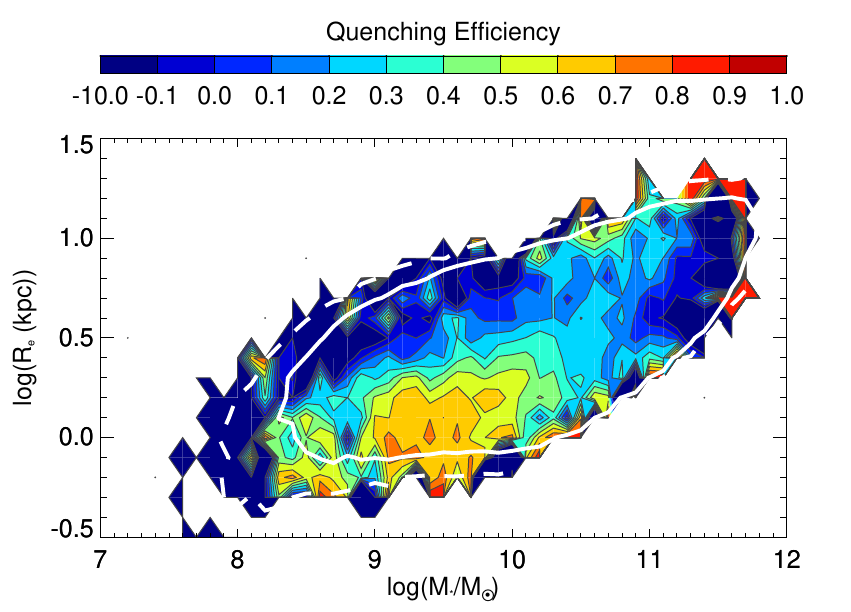}
}
	\caption{The quenching efficiency as a function of stellar mass M$_*$ and effective radius R$_e$.  The lowest contour includes everything below -0.1.}
\label{fig:qeff}
\end{figure}

The right panel of Figure~\ref{fig:csqf} shows the quiescent fraction of satellite galaxies.  Here the contours show more structure and are generally flatter; the quiescent fraction depends almost entirely on R$_e$ at $8.5<\log({M_*/M_\Sun})<9.5$ and increasingly more on mass at higher stellar masses.  To quantitatively compare these two populations, we show the difference in quenched fraction between satellites and centrals in Figure~\ref{fig:qfd}.  At almost all locations within this plane the difference is positive, indicating that satellites are more likely to be quenched than central galaxies of the same mass and size.  The difference ranges from 0 to $\sim 0.5$, with a broad maximum near M$_*$ $=$ 10$^{9.2}$ M$_\Sun$ and R$_e<2$ kpc.

To isolate the influence of environment from internal mechanisms, we now consider the quenching efficiency as a function of $M_*$ and $R_e$, in Figure~\ref{fig:qeff}.  The contours here show quite a simple structure, with a localized peak at M$_*$ $=$ 10$^{9.4}$ M$_\Sun$ and R$_e$ $=$ 1 kpc.  It is remarkable that this peak in efficiency is in the same location as the peak in satellite fraction, shown in Figure ~\ref{fig:fsat}.  This does not appear to be a trivial coincidence; we examine it 
further in Section~\ref{scn:IED}.

\begin{figure}
\centerline{
\includegraphics[clip=true,trim=0mm 0mm 0mm 0mm,scale=1,angle=0]{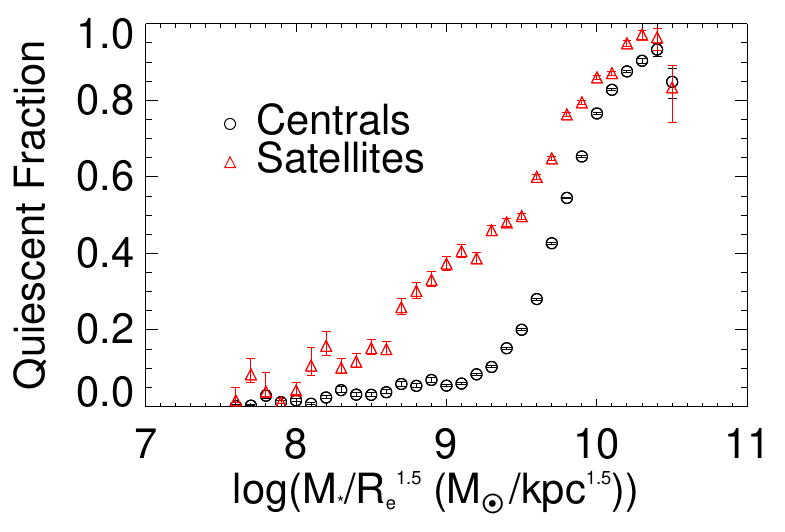}
}
	\caption{The quiescent fraction of satellite and central galaxies  as a function of M$_*$/R$_e^{1.5}$. Central galaxies are shown in black and satellite galaxies are in red. }
\label{fig:mr15_qf}
\end{figure}
\begin{figure}
\centerline{
		\includegraphics[clip=true,trim=0mm 0mm 0mm 0mm,scale=1,angle=0]{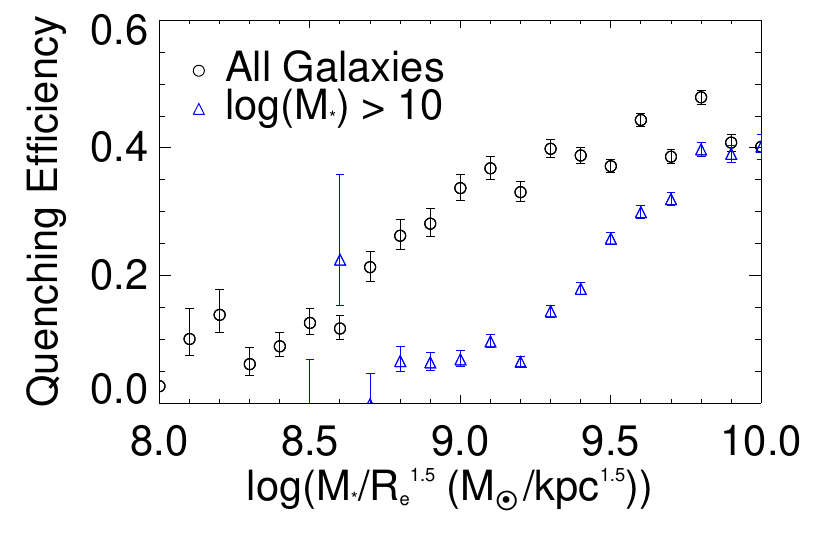}
}
	\caption{The quenching efficiency  as a function of M$_*$/R$_e^{1.5}$.  All galaxies are shown as black circles, while the subsample restricted to galaxies with M$_* > 10$ are shown as blue triangles.}
\label{fig:mr15_qeff}
\end{figure}
For most central galaxies (with $9<\log({M_*/M_\Sun})<11$), the shape of the quiescent fraction contours in Figure~\ref{fig:csqf} indicate that M$_*$/R$_e^{1.5}$ should be a more discerning parameter than either inferred velocity dispersion or surface density.  To show this explicitly, we plot the quiescent fraction as a function of M$_*$/R$_e^{1.5}$ in Figure~\ref{fig:mr15_qf}.  This should be compared directly with analogous plots as a function of $M_*$, $M_*/R_e$ and $M_*/R_e^2$ from Figure~\ref{fig:qfsis}. The fraction of quenched central galaxies increases very steeply, from $0.2$ to $0.8$ over a range of only $\sim0.5$ dex in M$_*$/R$_e^{1.5}$; recall that the same increase occurs over $\sim 0.7$ dex as a function of surface density and inferred velocity dispersion.  This intermediate dependence has appeared in the work of other authors that have considered related properties of galaxies \citep[e.g.][]{Franx,Cappellari,Barro}.  

Figure~\ref{fig:mr15_qf} also shows that the quenched fraction of satellite galaxies increases gradually as a function of M$_*$/R$_e^{1.5}$; it does not show the sharp threshold that is observed for central galaxies. 
The corresponding quenching efficiency as a function of M$_*$/R$_e^{1.5}$ is shown in Figure~\ref{fig:mr15_qeff}.  This also shows the same qualitative behaviour as it does as a function of the other quantities (c.f. Figure~\ref{fig:qe1d}), remaining relatively constant at $\sim 0.4$, above a certain threshold in each parameter. This reflects the fact that no one-dimensional quantity captures the complex dependence seen in Figure~\ref{fig:qeff}.

\section{Discussion}

\subsection{Quiescent Central Galaxies}\label{scn:discuss-centrals}
Comparing the two panels of Figure~\ref{fig:wdc}, it is evident that the observed trends in quiescent fraction are driven by the fact that the $M_*$--$R_e$ relation is different for star-forming and quiescent galaxies.  In fact, the central quiescent population shows two well-known sequences itself: a high-surface brightness population commonly identified with elliptical galaxies, and a lower-surface brightness population of ``spheroidal galaxies'' that is much more prominent among satellite galaxies.
The lack of evolution in the star--forming mass function, and corresponding growth in the quiescent galaxy mass function, makes it inevitable that galaxies must be migrating from the active population into the quiescent.  How, then, do these different structural relationships arise?  There are two general possibilities, that we will explore in turn.

\subsubsection{Hypothesis 1: Quenching probability depends on galaxy structure}\label{sec-hyp1}

First, we can consider that star-forming galaxies populate the $M_*$--$R_e$ plane as shown in the top panel of Figure~\ref{fig:wdc}, and that the probability of quenching is given by a function that is similar to\footnote{If the $M_*$--$R_e$ distribution of active galaxies evolves with time \citep[e.g.][]{Trujillo+06,Bruce+12}, the relevant quenching probability will not be identical to the quiescent fraction observed today.  Qualitatively, however, the scenario we describe still applies.} the contour shapes of Figure~\ref{fig:csqf}; in other words, it increases steadily with increasing $\sim M_*/R_e^{1.5}$. If this is the case, we might hope to find some other property of the star--forming population that depends on this quantity.  

\begin{figure}
\centerline{\includegraphics[clip=true,trim=0mm 0mm 0mm 0mm,scale=1.,angle=0]{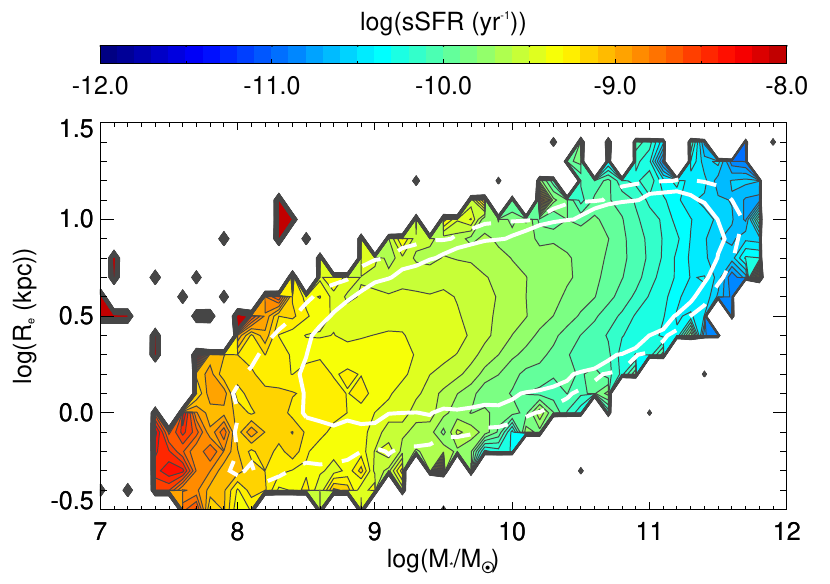}}
	\caption{The specific SFR of active, central galaxies are shown as coloured contours, as a function of $M_*$ and $R_e$.  For $M_*<10^{11}M_\Sun$, the sSFR shows a dependence on both $M_*$ and $R_e$, which is qualitatively similar to that of the quiescent fraction shown in Figure~\ref{fig:csqf}.
}
\label{fig:ssfrmr}
\end{figure}

For example, if the quenching process is a gradual one, we might see that amongst active galaxies, sSFR shows a similar dependence on $M_*$ and $R_e$ as the quiescent fraction of the whole sample.  This relation is shown as contours in Figure~\ref{fig:ssfrmr} and, indeed, the trends are qualitatively similar to the trends in quiescent fraction.  Specifically, for $M_*>10^{11}M_\Sun$ or so the contours are vertical, indicating the sSFR depends only on stellar mass, and is in fact quite low, representing a doubling time greater than a Hubble time.  For lower masses, however, the contours show a strong and non-trivial dependence on $R_e$.  For example, at fixed $M_*\sim10^{9.5}M_\Sun$, the sSFR varies with $R_e$ by a factor $\sim 3$, actually reaching a maximum for $R_e\sim 3$kpc.  This behaviour does not exactly match that of the quiescent fraction, but it is qualitatively consistent with what would be expected if high surface mass density galaxies are more likely to be quenched, over a finite timescale.

\subsubsection{Hypothesis 2: Structural changes associated with quenching}
\begin{figure*}
\centerline{
		\includegraphics[clip=true,trim=0mm 0mm 0mm 0mm,scale=.8,angle=0]{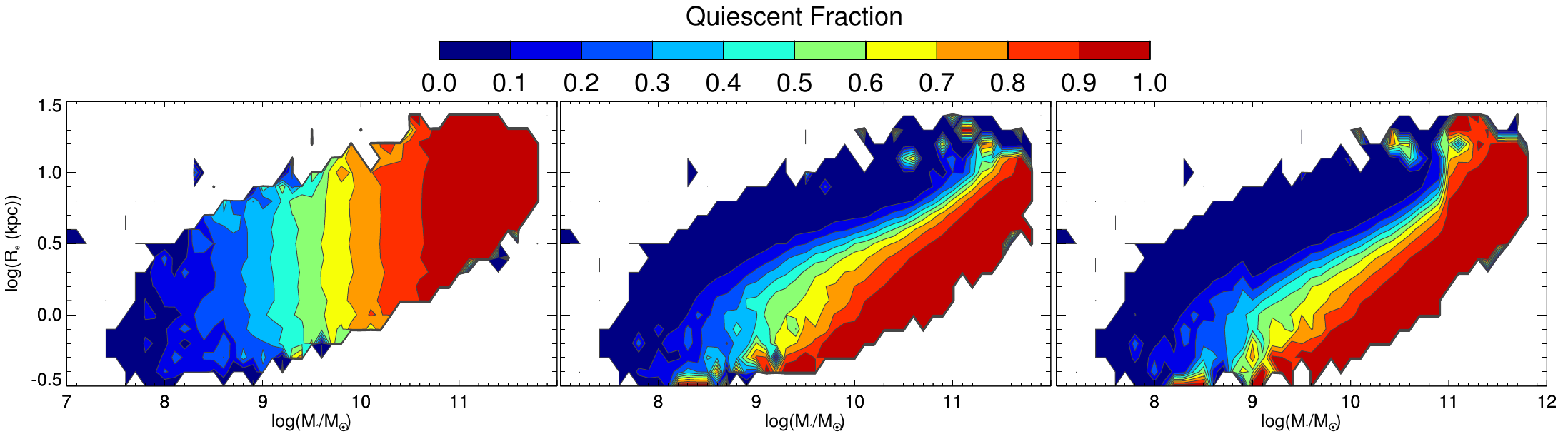}
}
	\caption{We show predictions of the quiescent fraction as a function of stellar mass M$_*$ and effective radius R$_e$ for three simple toy models, in which active galaxies taken from the distribution in the top panel of Figure~\ref{fig:wdc} are converted to quiescent galaxies, with a mass-dependent efficiency given by Equation~\ref{eqn:mqe}.  In the left panel, we assume that the quenching process results in no change in $R_e$; thus, the contours are vertical as expected.  In the middle panel, we assume that $\log({R_e})$ decreases by 0.25 when galaxies are quenched.  This provides a better match to the data in the left panel of Figure~\ref{fig:csqf}, but the predicted contours are too flat, and too widely spaced.  In the right panel, therefore, we show a model in which the amount by which $R_e$ decreases is mildly mass-dependent (smaller at higher masses; see text for details). This provides a reasonable, qualitative match to the data. }
\label{fig:mqqf}
\end{figure*}
An alternative interpretation of the contours in Figure~\ref{fig:csqf} is that
the quenching process occurs at a rate that depends only on $M_*$, but is accompanied by a change in structure.  This could occur if the disk component fades \citep[e.g.][]{Weinmann}, or is destroyed, or if the central density increases due to tidal torques inducing a central starburst, for example.  To roughly quantify this, we investigated a simple model in which some mass-dependent quenching process causes a galaxy's radius, but not mass, to decrease. 
We attempted to recreate the observed central quiescent fraction (Figure~\ref{fig:csqf}. \textit{left}) by arbitrarily quenching galaxies in the existing active central population, at a rate given by the mass-quenching efficiency of \citet{Peng10}, 
 \begin{equation}
QE_M=1-exp(-(M_*/p_1)^{p_2})
\label{eqn:mqe}
\end{equation}
with the parameters $p_1=10^{10.88} M_\Sun$ and $p_2=0.65$ adjusted to provide a good fit to our data, which are based on a different definition of ``quiescent''.  

To give a reference point, we first show the naive prediction that would result if quenching did not affect $R_e$, in the left panel of Figure~\ref{fig:mqqf}.  This recovers the expected result that the quiescent fraction does not depend on radius.  We then reduced the radius of all quenched galaxies by 0.25 dex, and the result is shown in the middle panel.  This does a reasonable job of qualitatively reproducing the data, but the contours are spaced too far apart, and are too flat at low masses, compared with the data in Figure~\ref{fig:csqf}.  In fact, the simplest interpretation of the quiescent fraction variation is that galaxies increase their stellar mass, as well as decrease $R_e$, upon quenching.  This is qualitatively consistent with the merging process, which likely plays a role especially at high masses \citep{Cappellari2}.  If, however, we retain the hypothesis that $M_*$ is unchanged, then we are driven to a model in which the decrease in $R_e$ is mass-dependent.  Such a model is shown in the right panel of Figure~\ref{fig:mqqf}, where we find reasonable qualitative agreement with the observations when $\log({R_e})$ is reduced by the discontinuous function
\begin{equation*}
\Delta\log{(R_e)}=
\begin{cases}
0.25  & \text{ if } \log{(M_*/M_\Sun)}<9\\
1.25-0.1\log(M_*/M_\Sun) & \text{ if } 9<\log(M_*/M_\Sun)<11\\
0.1 &\text{ if } \log(M_*/M_\Sun)>11,
\end{cases}
\end{equation*}
for $R_e$ measured in units of kpc.

This analysis simply illustrates that, if central galaxies are quenched at a rate that depends only on their stellar mass, this process must result in a reduction of $R_e$ by about a factor $\sim 2$, and perhaps a bit less for more massive galaxies.  It will be interesting to explore if more physical models, like those of \citet{Weinmann} based on disk-fading, for example, are consistent with these quantities.

\subsection{The role of the bulge}
The fact that the fraction of quiescent central galaxies depends strongly on surface density or inferred velocity dispersion makes it tempting to associate a quenching process with these very physical characteristics of the galaxy mass distribution.  However, Figure~\ref{fig:csqf} shows that these choices are simply projections of the more general dependence on $M_*$ and $R_e$, and that the best one-dimensional description of quiescent fraction based on these global quantities is actually $M_*/R_e^{1.5}$.  It is unclear from this analysis alone whether this quantity has physical significance, or if it is a secondary outcome of a dependence on something more fundamental, such as central velocity dispersion or the presence, or mass, of a central bulge.  This has been suggested by others \citep[e.g.][]{Martig,Fang,Cappellari2}, and could be a natural scenario if a central, supermassive black hole is responsible for the quenching, given the observed correlation between black hole masses and bulge properties \citep[e.g.][]{Magorrian,Gebhardt00,FM00}  The fact that the massive, quenched population is dominated by ``elliptical'' galaxies that likely formed at $z>2$ may also be good reason to consider the quenching associated with bulge-dominated galaxies to differ from that which leads to the buildup of the lower mass, ``spheroidal'' population.

The GIM2D fits to the surface brightness distributions provide information on the bulge mass, size and ellipticity; naturally these results are more sensitive to the models (and limitations of the data), than what we have considered so far.  For this analysis, the bulge component is identified from the two-component Gim2D fits to the $r-$band images, with the Sersic index fixed at $n=4$.  This is necessary because a free Sersic index results in many ``bulges'' with low-n, and it is not clear how best to treat them in this simple scenario.  

\begin{figure}
\centerline{\includegraphics[clip=true,trim=0mm 0mm 0mm 0mm,scale=1.,angle=0]{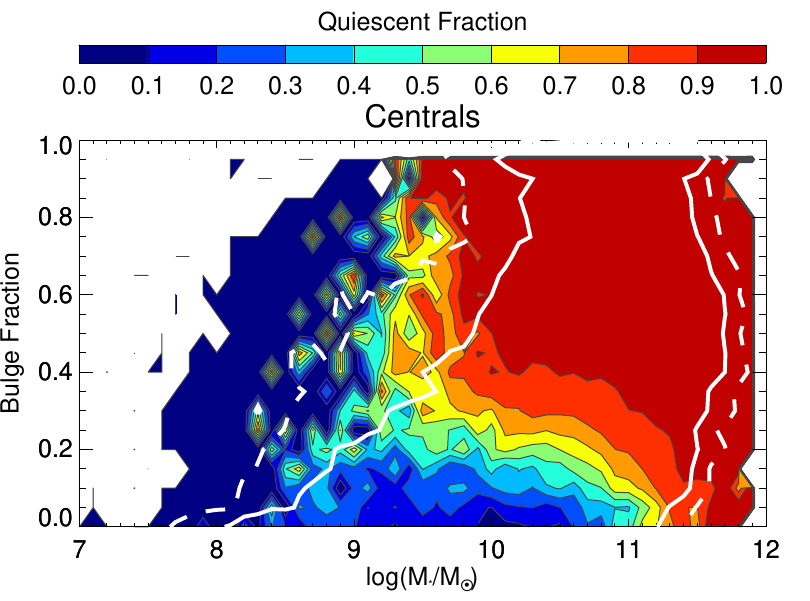}}
	\caption{Contours show the quiescent fraction of central galaxies, as a function of stellar mass and bulge fraction. This suggests two clear thresholds in stellar mass.  For $M_*<10^9M_\Sun$ almost all galaxies are forming stars, whether a bulge is present or not; conversely, for $M_*>10^{11}$ almost all galaxies are quiescent.  Between these masses, however, the fraction of quiescent galaxies appears to correlate with B/T in a way that is only weakly--dependent on mass.}
\label{fig:SMBT}
\end{figure}

\begin{figure}
\centerline{\includegraphics[clip=true,trim=0mm 0mm 0mm 0mm,scale=1.,angle=0]{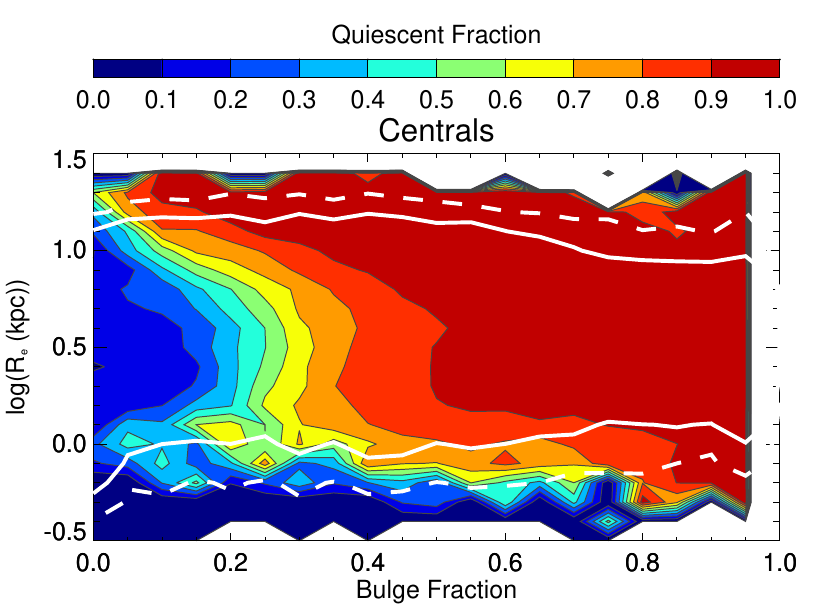}}
	\caption{Contours show the quiescent fraction of central galaxies, as a function of bulge fraction and $R_e$. Interestingly this shows that $R_e$ plays little role in establishing the quiescent fraction, independently from B/T.   The fraction of star forming galaxies decreases remarkably sharply, from $>90$ per cent for pure disk galaxies to $<20$ per cent for galaxies with as little as 40 per cent of their light in a bulge component.}
\label{fig:BTRe}
\end{figure}

We are first interested in how the quiescent fraction depends on three parameters: $M_*$, $R_e$ and the fraction of $r-$band luminosity contained in the bulge component, $B/T$.  We show this as contours of constant quiescent fraction as a function of $M_*$ and $B/T$ in Figure~\ref{fig:SMBT}, and as a function of $B/T$ and $R_e$ in Figure~\ref{fig:BTRe}.  These figures seem to indicate that the quiescent fraction is determined primarily from $M_*$ and $B/T$ in a largely separable way.  There are two thresholds in stellar mass.  For galaxies with $M_*<10^9M_\Sun$, all galaxies are forming stars, regardless of their $B/T$, while for the most massive galaxies ($M_*>10^{11.5}M_\Sun$), almost no galaxies are forming stars.  But between these limits, the quiescent fraction depends mostly on $B/T$, varying from $0$ to $0.8$ as B/T varies from $0$ to only $0.3$.  Figure~\ref{fig:BTRe} shows similar behaviour when $R_e$ is considered as a variable, with quiescent fraction depending only on $B/T$ over the range $0.6<R_e<10$ range that encompasses the great majority of galaxies in our sample.   

\begin{figure}
\centerline{\includegraphics[clip=true,trim=0mm 0mm 0mm 0mm,scale=1.,angle=0]{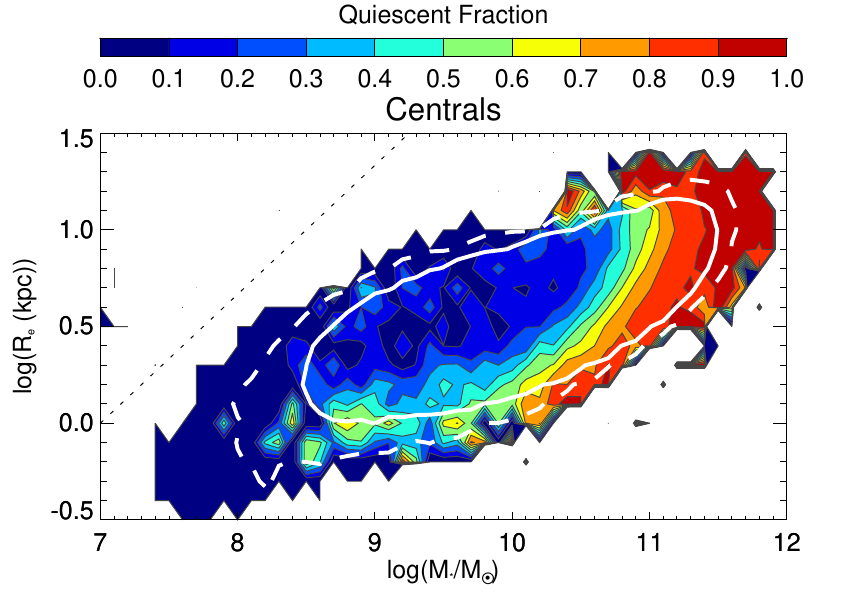}}
	\caption{We show the quiescent fraction of central galaxies, as in Figure~\ref{fig:csqf}, but for disk dominated (B/T $<0.3$) galaxies only. The shape of the contours in this plane is similar to that of the whole central galaxy population (left panel of Figure~\ref{fig:csqf}), demonstrating that the trend is not driven by bulge-dominated galaxies.}
\label{fig:BTlt03}
\end{figure}

It is apparent that galaxies in this sample with a substantial bulge, $B/T>0.3$ or so, are almost all quiescent.  This necessarily includes most of the high mass-density galaxies that we identify with ``ellipticals'' following \citet{KormendyBender}, and which dominate the $M_*-R_e$ distribution of quiescent galaxies in Figure~\ref{fig:wdc}.  It is interesting to consider any remaining trends in quiescent fraction if these galaxies are removed, and we show the result in Figure~\ref{fig:BTlt03}.  This is the equivalent of the left panel of Figure~\ref{fig:csqf}, restricted to disk-dominated galaxies, and we observe the same strong dependence on $M_*/R_e^{1.5}$.  Thus we interpret the trend as due, not to the varying mix of bulge-dominated and disk-dominated galaxies with $M_*$ and $R_e$, but as a result of the simple correlation shown in Figure~\ref{fig:SMBT} between quiescent fraction and bulge fraction amongst galaxies with both components.
It may be this correlation that is responsible for the more complex dependence on $M_*/R_e^{1.5}$; in any case it reinforces the fact that is not correct to assume galaxy mass is the driving parameter \citep[e.g.][]{Peng10}.  
If the bulge itself does play a role, it will be of interest to investigate these correlations more carefully, with an accurate determination of the bulge mass and structure \citep[e.g.][]{Mendel13}.  

\subsection{Satellite Quenching} \label{scn:IED}
\begin{figure}
\begin{center}
\includegraphics[clip=true,trim=0mm 0mm 0mm 0mm,scale=1,angle=0]{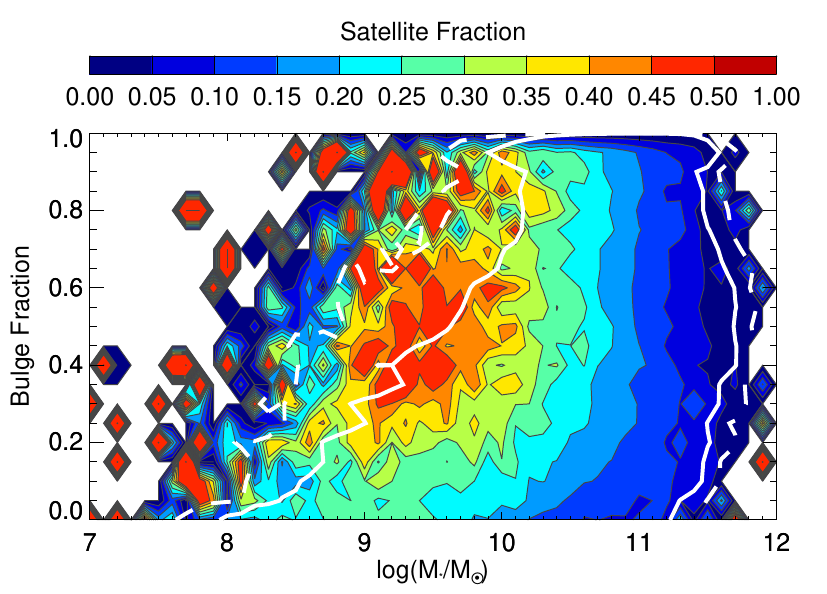}
\caption{The fraction of satellite galaxies is shown as a function of $M_*$ and bulge fraction.  Relative to central galaxies, satellites are more common at low masses, $9<M_*/M_\Sun<9.5$, and intermediate bulge fractions, $0.3<B/T<0.7$.}
\label{fig:fsatBT}
\end{center}
\end{figure}

\begin{figure}
\begin{center}
\includegraphics[clip=true,trim=0mm 0mm 0mm 0mm,scale=1,angle=0]{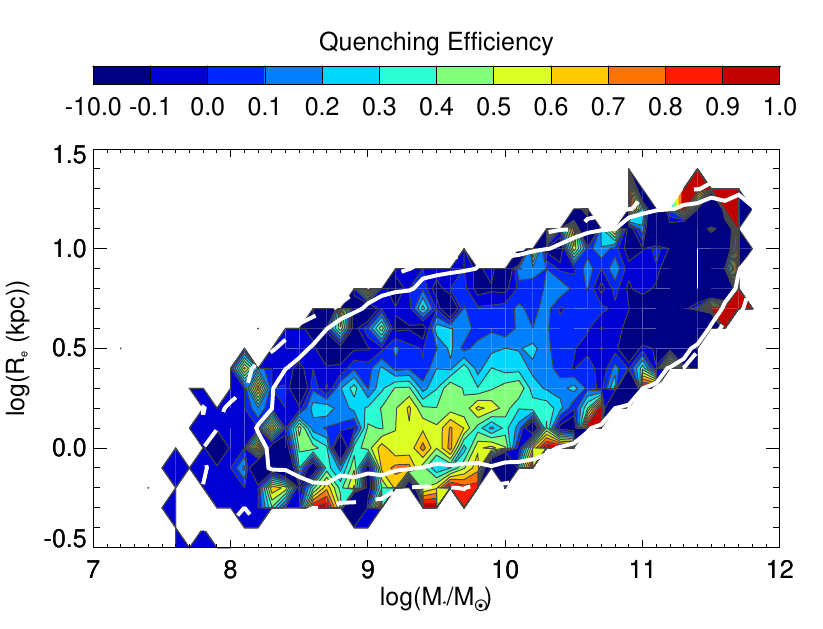}
\caption{The quenching efficiency is shown as a function of $M_*$ and $R_e$, for a sample in which the M$_*$-R$_e$-B/T distribution is forced to be the same for the central and satellite galaxies.  This still reaches a localized maximum of $\sim 0.7$, for galaxies with log$(M_*/M_\Sun)\sim 9.4$ and $R_e\sim 1$ kpc. }
\label{fig:qematched2d}
\end{center}
\end{figure}

Figure~\ref{fig:fsat} showed that satellite galaxies are distributed differently in $M_*$ and $R_e$ than central galaxies, and in Figure~\ref{fig:fsatBT} we show the well-known result that this difference in structure also extends to bulge fraction.  Interestingly, the low mass galaxies that dominate the satellite population are more likely to have intermediate bulge fractions, $0.3<B/T<0.7$.  Amongst central galaxies, such bulge-dominated galaxies are almost all quiescent.  Thus it is interesting to revisit the question of whether galaxies with identical morphological structure still show a dependence on whether they are a central or satellite galaxy.  We do this by creating a randomly matched sample of satellite galaxies, that has the same abundance as central galaxies at every $M_*$, $R_e$ and $B/T$.  We then recreate the quenching efficiency plot shown in Figure~\ref{fig:qeff}, but where the comparison is now made between populations that are structurally similar, in Figure~\ref{fig:qematched2d}.  The result is almost unchanged, and shows that quenching efficiency is almost always positive, and still reaches a localized maximum of $\sim 0.7$ at the same region of parameter space.

It is remarkable that Figure~\ref{fig:qematched2d} shows that satellite quenching efficiency is certainly not a function of stellar mass alone.  In fact, it reaches a maximum value in the same localized region where there is an excess of satellite galaxies, relative to the number of centrals (Figure~\ref{fig:fsat}).  It is possible that this is a coincidence, and satellite galaxies just happen to most commonly have a structure for which quenching efficiency is high. 
On the other hand, if we dismiss the possibility that this is a coincidence, it implies that satellite quenching {\it must} be accompanied by a change in structure; as star formation shuts down, galaxies migrate toward that region of $M_*,$ $R_e,$ $B/T$ space where most satellite galaxies are found.  In this case, quenching efficiency does not have the physical meaning that we ascribed to it, since satellite galaxies are not being compared with their rightful, central progenitors.  For satellites moreso than centrals, it may also be the case that the stellar mass changes during the quenching process; this would considerably complicate the analysis.

\section{Summary and Conclusions}

We have used a sample of 471,554 galaxies at 0.01 $<$ z $<$ 0.2 selected from the SDSS DR7, to examine how the fraction of quiescent galaxies depends upon stellar mass, effective radius, bulge fraction, and whether a galaxy is central or satellite in the group catalogue of \citet{YangDR7}.  We use the GIM2D parametric fits of \citet{Simard} to determine the circular half-light radius ($R_e$) of all galaxies in the sample, and the specific star formation rates and stellar masses of \citet{Brinchmann} to classify galaxies as active (on the star--forming main sequence) or quiescent.  We make the following observations:

\begin{enumerate}

\item The fraction of quiescent galaxies is a steeper function of M$_*$/R$_e$ and M$_*$/R$_e^2$ than it is of stellar mass alone.  This confirms the findings of many other authors
\citep[e.g.][]{Kauffmann2,Franx,Wake2,Wake1,Cheung}.

\item The quiescent fraction of satellite galaxies is always higher than that of central galaxies, except perhaps at the highest masses where almost all galaxies are quenched centrals.  Thus a galaxy's status as satellite or central in its halo plays a role in its evolution.  The quenching efficiency \citep{Peng10}, which attempts to quantify the fraction of satellite galaxies that had their star formation quenched as a satellite, is $\sim 40$ per cent, approximately independent of stellar mass for $M_*>10^9M_\Sun$.

\item We presented the quiescent fraction of central galaxies as function of both $M_\ast$ and $R_e$ in Figure~\ref{fig:qfd}.  This representation makes it clear that the fraction does not depend on stellar mass alone, and that over most of the parameter space it appears to depend most strongly on $\sim$M$_*$/R$_e^{1.5}$.  

\item We showed that the quiescent fraction of central galaxies correlates strongly with $B/T$ in a simple way that largely accounts for the more complex dependence on $R_e$. Galaxies with more than 40 per cent of their light in a bulge component are almost all quiescent, and galaxies less massive than $M_*=10^9M_\Sun$ are almost all forming stars.  Otherwise, the quiescent fraction increases monotonically with increasing B/T at fixed $M_*$. 

\item The quenching efficiency is highest for galaxies with 10$^{9}<$ M$_*<$ 10$^{10}$ M$_\Sun$ and 10$^{-0.2}<$ R$_e<$ 10$^{0.2}$ kpc (Figure~\ref{fig:qeff}), but appears roughly constant above a certain threshold when taken solely as a function of M$_*$, M$_*$/R$_e$, or M$_*$/R$_e^2$.

\item There are structural differences between satellite and central galaxies, such that, at a given stellar mass, satellite galaxies tend to have smaller $R_e$ and more light in a bulge component.  The fraction of satellites is greatest (at about 50 per cent) for galaxies with $9.0<\log({M_*}/M_\Sun)<10.0$, and $R_e<2$kpc.    Even after accounting for this difference by matching central and satellite galaxies to have the same $M_*$, $R_e$, and B/T, the quenching efficiency remains positive at all masses, and reaches a maximum of 70 per cent for galaxies with $\log(M_*/M_\Sun)\sim 9.4$ and $R_e\sim$1 kpc.

\end{enumerate}

For central galaxies, it is appealing to consider a probability of star formation activity ending that depends on their internal structure.  In particular the strong correlation between quiescent fraction and  $B/T$ points to the dominance of a central bulge component as the driving parameter.  
We showed that, amongst star--forming central galaxies with $M_*<10^{11}M_\Sun$, the average sSFR decreases with increasing $M_*/R_e$, qualitatively suporting a picture in which the presence of a bulge  reduces the sSFR and increases the probability of being quenched.  Alternatively, the data could be explained if the probability of quenching depends only on $M_*$, but is 
accompanied by a change in structure, with $R_e$ shrinking by a factor $\sim 2$ due to the increased prominence of a bulge when star formation is terminated.

For satellite galaxies, that fact that the quenching efficiency is a maximum for the same combination of $M_*$ and $R_e$ at which the satellite fraction itself peaks almost certainly means that their $R_e$, and perhaps their mass, changes upon quenching.  

\section{Acknowledgments}
CMBO acknowledges support from an NSERC Undergraduate Student Research Award.  MLB acknowledges support from an NSERC Discovery Grant, and NOVA and NWO grants that supported his sabbatical leave at the Sterrewacht at Leiden University, where this work was completed.  MLB thanks Sean McGee, Asa Bluck, Sara Ellison and an anonymous referee for their careful reading of the manuscript, and for helpful comments which improved our interpretation of the data.

Funding for the SDSS and SDSS-II has been provided by the Alfred P. Sloan Foundation, the Participating Institutions, the National Science Foundation, the U.S. Department of Energy, the National Aeronautics and Space Administration, the Japanese Monbukagakusho, the Max Planck Society, and the Higher Education Funding Council for England. The SDSS Web Site is http://www.sdss.org/.  The SDSS is managed by the Astrophysical Research Consortium for the Participating Institutions. The Participating Institutions are the American Museum of Natural History, Astrophysical Institute Potsdam, University of Basel, University of Cambridge, Case Western Reserve University, University of Chicago, Drexel University, Fermilab, the Institute for Advanced Study, the Japan Participation Group, Johns Hopkins University, the Joint Institute for Nuclear Astrophysics, the Kavli Institute for Particle Astrophysics and Cosmology, the Korean Scientist Group, the Chinese Academy of Sciences (LAMOST), Los Alamos National Laboratory, the Max-Planck-Institute for Astronomy (MPIA), the Max-Planck-Institute for Astrophysics (MPA), New Mexico State University, Ohio State University, University of Pittsburgh, University of Portsmouth, Princeton University, the United States Naval Observatory, and the University of Washington. 

\bibliography{ms}

\begin{thebibliography}{87}
\expandafter\ifx\csname natexlab\endcsname\relax\def\natexlab#1{#1}\fi

\bibitem[{{Abazajian} {et~al}\mbox{.}(2009){Abazajian}, {Adelman-McCarthy},
  {Ag{\"u}eros}, {Allam}, {Allende Prieto}, {An}, {Anderson}, {Anderson},
  {Annis}, {Bahcall}, \& et~al.}]{DR7}
{Abazajian} K.~N. {et~al.}, 2009, \apjs, 182, 543

\bibitem[{{Allen} {et~al}\mbox{.}(2006){Allen}, {Driver}, {Graham}, {Cameron},
  {Liske}, \& {de Propris}}]{Allen}
{Allen} P.~D., {Driver} S.~P., {Graham} A.~W., {Cameron} E., {Liske} J., {de
  Propris} R., 2006, \mnras, 371, 2

\bibitem[{{Baldry} {et~al}\mbox{.}(2006){Baldry}, {Balogh}, {Bower},
  {Glazebrook}, {Nichol}, {Bamford}, \& {Budavari}}]{Baldry2}
{Baldry} I.~K., {Balogh} M.~L., {Bower} R.~G., {Glazebrook} K., {Nichol} R.~C.,
  {Bamford} S.~P., {Budavari} T., 2006, \mnras, 373, 469

\bibitem[{{Baldry} {et~al}\mbox{.}(2004){Baldry}, {Glazebrook}, {Brinkmann},
  {Ivezi{\'c}}, {Lupton}, {Nichol}, \& {Szalay}}]{Baldry}
{Baldry} I.~K., {Glazebrook} K., {Brinkmann} J., {Ivezi{\'c}} {\v Z}., {Lupton}
  R.~H., {Nichol} R.~C., {Szalay} A.~S., 2004, \apj, 600, 681

\bibitem[{{Balogh} {et~al}\mbox{.}(2004){Balogh}, {Baldry}, {Nichol}, {Miller},
  {Bower}, \& {Glazebrook}}]{Balogh2}
{Balogh} M.~L., {Baldry} I.~K., {Nichol} R., {Miller} C., {Bower} R.,
  {Glazebrook} K., 2004, \apjl, 615, L101

\bibitem[{{Balogh} {et~al}\mbox{.}(2000){Balogh}, {Navarro}, \&
  {Morris}}]{Balogh1}
{Balogh} M.~L., {Navarro} J.~F., {Morris} S.~L., 2000, \apj, 540, 113

\bibitem[{{Barro} {et~al}\mbox{.}(2013){Barro}, {Faber},
  {P{\'e}rez-Gonz{\'a}lez}, {Koo}, {Williams}, {Kocevski}, {Trump}, {Mozena},
  {McGrath}, {van der Wel}, {Wuyts}, {Bell}, {Croton}, {Ceverino}, {Dekel},
  {Ashby}, {Cheung}, {Ferguson}, {Fontana}, {Fang}, {Giavalisco}, {Grogin},
  {Guo}, {Hathi}, {Hopkins}, {Huang}, {Koekemoer}, {Kartaltepe}, {Lee},
  {Newman}, {Porter}, {Primack}, {Ryan}, {Rosario}, {Somerville}, {Salvato}, \&
  {Hsu}}]{Barro}
{Barro} G. {et~al.}, 2013, \apj, 765, 104

\bibitem[{{Bell}(2008)}]{Bell2}
{Bell} E.~F., 2008, \apj, 682, 355

\bibitem[{{Bell} {et~al}\mbox{.}(2004){Bell}, {Wolf}, {Meisenheimer}, {Rix},
  {Borch}, {Dye}, {Kleinheinrich}, {Wisotzki}, \& {McIntosh}}]{Bell}
{Bell} E.~F. {et~al.}, 2004, \apj, 608, 752

\bibitem[{{Bennett} {et~al}\mbox{.}(2013){Bennett}, {Larson}, {Weiland},
  {Jarosik}, {Hinshaw}, {Odegard}, {Smith}, {Hill}, {Gold}, {Halpern},
  {Komatsu}, {Nolta}, {Page}, {Spergel}, {Wollack}, {Dunkley}, {Kogut},
  {Limon}, {Meyer}, {Tucker}, \& {Wright}}]{WMAP}
{Bennett} C.~L. {et~al.}, 2013, \apjs, 208, 20

\bibitem[{{Berlind} {et~al}\mbox{.}(2005){Berlind}, {Blanton}, {Hogg},
  {Weinberg}, {Dav{\'e}}, {Eisenstein}, \& {Katz}}]{Berlind}
{Berlind} A.~A., {Blanton} M.~R., {Hogg} D.~W., {Weinberg} D.~H., {Dav{\'e}}
  R., {Eisenstein} D.~J., {Katz} N., 2005, \apj, 629, 625

\bibitem[{{Blanton} {et~al}\mbox{.}(2003{\natexlab{a}}){Blanton}, {Brinkmann},
  {Csabai}, {Doi}, {Eisenstein}, {Fukugita}, {Gunn}, {Hogg}, \&
  {Schlegel}}]{Blanton2003a}
{Blanton} M.~R. {et~al.}, 2003{\natexlab{a}}, \aj, 125, 2348

\bibitem[{{Blanton} {et~al}\mbox{.}(2003{\natexlab{b}}){Blanton}, {Hogg},
  {Bahcall}, {Baldry}, {Brinkmann}, {Csabai}, {Eisenstein}, {Fukugita}, {Gunn},
  {Ivezi{\'c}}, {Lamb}, {Lupton}, {Loveday}, {Munn}, {Nichol}, {Okamura},
  {Schlegel}, {Shimasaku}, {Strauss}, {Vogeley}, \& {Weinberg}}]{Blanton2003c}
{Blanton} M.~R. {et~al.}, 2003{\natexlab{b}}, \apj, 594, 186

\bibitem[{{Blanton} {et~al}\mbox{.}(2003{\natexlab{c}}){Blanton}, {Hogg},
  {Bahcall}, {Brinkmann}, {Britton}, {Connolly}, {Csabai}, {Fukugita},
  {Loveday}, {Meiksin}, {Munn}, {Nichol}, {Okamura}, {Quinn}, {Schneider},
  {Shimasaku}, {Strauss}, {Tegmark}, {Vogeley}, \& {Weinberg}}]{Blanton2003b}
{Blanton} M.~R. {et~al.}, 2003{\natexlab{c}}, \apj, 592, 819

\bibitem[{{Blanton} \& {Roweis}(2007)}]{BlantonRoweis}
{Blanton} M.~R., {Roweis} S., 2007, \aj, 133, 734

\bibitem[{{Blanton} {et~al}\mbox{.}(2005){Blanton}, {Schlegel}, {Strauss},
  {Brinkmann}, {Finkbeiner}, {Fukugita}, {Gunn}, {Hogg}, {Ivezi{\'c}}, {Knapp},
  {Lupton}, {Munn}, {Schneider}, {Tegmark}, \& {Zehavi}}]{Blanton2005}
{Blanton} M.~R. {et~al.}, 2005, \aj, 129, 2562

\bibitem[{{Brammer} {et~al}\mbox{.}(2009){Brammer}, {Whitaker}, {van Dokkum},
  {Marchesini}, {Labb{\'e}}, {Franx}, {Kriek}, {Quadri}, {Illingworth}, {Lee},
  {Muzzin}, \& {Rudnick}}]{Brammer}
{Brammer} G.~B. {et~al.}, 2009, \apjl, 706, L173

\bibitem[{{Brinchmann} {et~al}\mbox{.}(2004){Brinchmann}, {Charlot}, {White},
  {Tremonti}, {Kauffmann}, {Heckman}, \& {Brinkmann}}]{Brinchmann}
{Brinchmann} J., {Charlot} S., {White} S.~D.~M., {Tremonti} C., {Kauffmann} G.,
  {Heckman} T., {Brinkmann} J., 2004, \mnras, 351, 1151

\bibitem[{{Brown} {et~al}\mbox{.}(2007){Brown}, {Dey}, {Jannuzi}, {Brand},
  {Benson}, {Brodwin}, {Croton}, \& {Eisenhardt}}]{Brown}
{Brown} M.~J.~I., {Dey} A., {Jannuzi} B.~T., {Brand} K., {Benson} A.~J.,
  {Brodwin} M., {Croton} D.~J., {Eisenhardt} P.~R., 2007, \apj, 654, 858

\bibitem[{{Bruce} {et~al}\mbox{.}(2012){Bruce}, {Dunlop}, {Cirasuolo},
  {McLure}, {Targett}, {Bell}, {Croton}, {Dekel}, {Faber}, {Ferguson},
  {Grogin}, {Kocevski}, {Koekemoer}, {Koo}, {Lai}, {Lotz}, {McGrath}, {Newman},
  \& {van der Wel}}]{Bruce+12}
{Bruce} V.~A. {et~al.}, 2012, \mnras, 427, 1666

\bibitem[{{Cameron}(2011)}]{Cameron}
{Cameron} E., 2011, \pasa, 28, 128

\bibitem[{{Cappellari}(2013)}]{Cappellari2}
{Cappellari} M., 2013, ArXiv e-prints

\bibitem[{{Cappellari} {et~al}\mbox{.}(2013){Cappellari}, {McDermid},
  {Alatalo}, {Blitz}, {Bois}, {Bournaud}, {Bureau}, {Crocker}, {Davies},
  {Davis}, {de Zeeuw}, {Duc}, {Emsellem}, {Khochfar}, {Krajnovi{\'c}},
  {Kuntschner}, {Morganti}, {Naab}, {Oosterloo}, {Sarzi}, {Scott}, {Serra},
  {Weijmans}, \& {Young}}]{Cappellari}
{Cappellari} M. {et~al.}, 2013, \mnras, 432, 1862

\bibitem[{{Carter} {et~al}\mbox{.}(2001){Carter}, {Fabricant}, {Geller},
  {Kurtz}, \& {McLean}}]{Carter}
{Carter} B.~J., {Fabricant} D.~G., {Geller} M.~J., {Kurtz} M.~J., {McLean} B.,
  2001, \apj, 559, 606

\bibitem[{{Cheung} {et~al}\mbox{.}(2012){Cheung}, {Faber}, {Koo}, {Dutton},
  {Simard}, {McGrath}, {Huang}, {Bell}, {Dekel}, {Fang}, {Salim}, {Barro},
  {Bundy}, {Coil}, {Cooper}, {Conselice}, {Davis}, {Dom{\'{\i}}nguez},
  {Kassin}, {Kocevski}, {Koekemoer}, {Lin}, {Lotz}, {Newman}, {Phillips},
  {Rosario}, {Weiner}, \& {Willmer}}]{Cheung}
{Cheung} E. {et~al.}, 2012, \apj, 760, 131

\bibitem[{{Davis} \& {Geller}(1976)}]{DavisGeller}
{Davis} M., {Geller} M.~J., 1976, \apj, 208, 13

\bibitem[{{D'Onofrio} {et~al}\mbox{.}(2011){D'Onofrio}, {Valentinuzzi},
  {Fasano}, {Moretti}, {Bettoni}, {Poggianti}, {Vulcani}, {Varela}, {Fritz},
  {Cava}, {Kj{\ae}rgaard}, {Moles}, {Couch}, \& {Dressler}}]{Donofrio+11}
{D'Onofrio} M. {et~al.}, 2011, \apjl, 727, L6

\bibitem[{{Dressler}(1980)}]{Dressler}
{Dressler} A., 1980, \apj, 236, 351

\bibitem[{{Driver} {et~al}\mbox{.}(2006){Driver}, {Allen}, {Graham}, {Cameron},
  {Liske}, {Ellis}, {Cross}, {De Propris}, {Phillipps}, \& {Couch}}]{Driver}
{Driver} S.~P. {et~al.}, 2006, \mnras, 368, 414

\bibitem[{{Faber} {et~al}\mbox{.}(2007){Faber}, {Willmer}, {Wolf}, {Koo},
  {Weiner}, {Newman}, {Im}, {Coil}, {Conroy}, {Cooper}, {Davis}, {Finkbeiner},
  {Gerke}, {Gebhardt}, {Groth}, {Guhathakurta}, {Harker}, {Kaiser}, {Kassin},
  {Kleinheinrich}, {Konidaris}, {Kron}, {Lin}, {Luppino}, {Madgwick},
  {Meisenheimer}, {Noeske}, {Phillips}, {Sarajedini}, {Schiavon}, {Simard},
  {Szalay}, {Vogt}, \& {Yan}}]{Faber}
{Faber} S.~M. {et~al.}, 2007, \apj, 665, 265

\bibitem[{{Fang} {et~al}\mbox{.}(2013){Fang}, {Faber}, {Koo}, \&
  {Dekel}}]{Fang}
{Fang} J.~J., {Faber} S.~M., {Koo} D.~C., {Dekel} A., 2013, \apj, 776, 63

\bibitem[{{Ferrarese} \& {Merritt}(2000)}]{FM00}
{Ferrarese} L., {Merritt} D., 2000, \apjl, 539, L9

\bibitem[{{Franx} {et~al}\mbox{.}(2008){Franx}, {van Dokkum}, {Schreiber},
  {Wuyts}, {Labb{\'e}}, \& {Toft}}]{Franx}
{Franx} M., {van Dokkum} P.~G., {Schreiber} N.~M.~F., {Wuyts} S., {Labb{\'e}}
  I., {Toft} S., 2008, \apj, 688, 770

\bibitem[{{Gebhardt} {et~al}\mbox{.}(2000){Gebhardt}, {Bender}, {Bower},
  {Dressler}, {Faber}, {Filippenko}, {Green}, {Grillmair}, {Ho}, {Kormendy},
  {Lauer}, {Magorrian}, {Pinkney}, {Richstone}, \& {Tremaine}}]{Gebhardt00}
{Gebhardt} K. {et~al.}, 2000, \apjl, 539, L13

\bibitem[{{Geha} {et~al}\mbox{.}(2012){Geha}, {Blanton}, {Yan}, \&
  {Tinker}}]{Geha}
{Geha} M., {Blanton} M.~R., {Yan} R., {Tinker} J.~L., 2012, \apj, 757, 85

\bibitem[{{Gilbank} {et~al}\mbox{.}(2010){Gilbank}, {Baldry}, {Balogh},
  {Glazebrook}, \& {Bower}}]{Gilbank}
{Gilbank} D.~G., {Baldry} I.~K., {Balogh} M.~L., {Glazebrook} K., {Bower}
  R.~G., 2010, \mnras, 405, 2594

\bibitem[{{Gon{\c c}alves} {et~al}\mbox{.}(2012){Gon{\c c}alves}, {Martin},
  {Men{\'e}ndez-Delmestre}, {Wyder}, \& {Koekemoer}}]{Goncalves}
{Gon{\c c}alves} T.~S., {Martin} D.~C., {Men{\'e}ndez-Delmestre} K., {Wyder}
  T.~K., {Koekemoer} A., 2012, \apj, 759, 67

\bibitem[{{Grebel} {et~al}\mbox{.}(2003){Grebel}, {Gallagher}, \&
  {Harbeck}}]{Grebel}
{Grebel} E.~K., {Gallagher}, III J.~S., {Harbeck} D., 2003, \aj, 125, 1926

\bibitem[{{Gunn} \& {Gott}(1972)}]{Gunn}
{Gunn} J.~E., {Gott}, III J.~R., 1972, \apj, 176, 1

\bibitem[{{Hogg} {et~al}\mbox{.}(2003){Hogg}, {Blanton}, {Eisenstein}, {Gunn},
  {Schlegel}, {Zehavi}, {Bahcall}, {Brinkmann}, {Csabai}, {Schneider},
  {Weinberg}, \& {York}}]{Hogg}
{Hogg} D.~W. {et~al.}, 2003, \apjl, 585, L5

\bibitem[{{Hoyos} {et~al}\mbox{.}(2011){Hoyos}, {den Brok}, {Verdoes Kleijn},
  {Carter}, {Balcells}, {Guzm{\'a}n}, {Peletier}, {Ferguson}, {Goudfrooij},
  {Graham}, {Hammer}, {Karick}, {Lucey}, {Matkovi{\'c}}, {Merritt}, {Mouhcine},
  \& {Valentijn}}]{Hoyos}
{Hoyos} C. {et~al.}, 2011, \mnras, 411, 2439

\bibitem[{{Kauffmann} {et~al}\mbox{.}(2006){Kauffmann}, {Heckman}, {De Lucia},
  {Brinchmann}, {Charlot}, {Tremonti}, {White}, \& {Brinkmann}}]{Kauffmann2}
{Kauffmann} G., {Heckman} T.~M., {De Lucia} G., {Brinchmann} J., {Charlot} S.,
  {Tremonti} C., {White} S.~D.~M., {Brinkmann} J., 2006, \mnras, 367, 1394

\bibitem[{{Kauffmann} {et~al}\mbox{.}(2003){Kauffmann}, {Heckman}, {White},
  {Charlot}, {Tremonti}, {Brinchmann}, {Bruzual}, {Peng}, {Seibert},
  {Bernardi}, {Blanton}, {Brinkmann}, {Castander}, {Cs{\'a}bai}, {Fukugita},
  {Ivezic}, {Munn}, {Nichol}, {Padmanabhan}, {Thakar}, {Weinberg}, \&
  {York}}]{Kauffmann3}
{Kauffmann} G. {et~al.}, 2003, \mnras, 341, 33

\bibitem[{{Kawata} \& {Mulchaey}(2008)}]{Kawata}
{Kawata} D., {Mulchaey} J.~S., 2008, \apjl, 672, L103

\bibitem[{{Kormendy}(1985)}]{Kormendy85}
{Kormendy} J., 1985, \apj, 295, 73

\bibitem[{{Kormendy} \& {Bender}(2012)}]{KormendyBender}
{Kormendy} J., {Bender} R., 2012, \apjs, 198, 2

\bibitem[{{Kormendy} {et~al}\mbox{.}(2009){Kormendy}, {Fisher}, {Cornell}, \&
  {Bender}}]{Kormendy}
{Kormendy} J., {Fisher} D.~B., {Cornell} M.~E., {Bender} R., 2009, \apjs, 182,
  216

\bibitem[{{Kroupa}(2001)}]{Kroupa}
{Kroupa} P., 2001, \mnras, 322, 231

\bibitem[{{Magorrian} {et~al}\mbox{.}(1998){Magorrian}, {Tremaine},
  {Richstone}, {Bender}, {Bower}, {Dressler}, {Faber}, {Gebhardt}, {Green},
  {Grillmair}, {Kormendy}, \& {Lauer}}]{Magorrian}
{Magorrian} J. {et~al.}, 1998, \aj, 115, 2285

\bibitem[{{Martig} {et~al}\mbox{.}(2009){Martig}, {Bournaud}, {Teyssier}, \&
  {Dekel}}]{Martig}
{Martig} M., {Bournaud} F., {Teyssier} R., {Dekel} A., 2009, \apj, 707, 250

\bibitem[{{McGee} {et~al}\mbox{.}(2011){McGee}, {Balogh}, {Wilman}, {Bower},
  {Mulchaey}, {Parker}, \& {Oemler}}]{MBW+}
{McGee} S.~L., {Balogh} M.~L., {Wilman} D.~J., {Bower} R.~G., {Mulchaey} J.~S.,
  {Parker} L.~C., {Oemler} A., 2011, \mnras, 413, 996

\bibitem[{{Mendel} {et~al}\mbox{.}(2013){Mendel}, {Simard}, {Palmer},
  {Ellison}, \& {Patton}}]{Mendel13}
{Mendel} J.~T., {Simard} L., {Palmer} M., {Ellison} S.~L., {Patton} D.~R.,
  2013, ArXiv e-prints

\bibitem[{{Moore} {et~al}\mbox{.}(1996){Moore}, {Katz}, {Lake}, {Dressler}, \&
  {Oemler}}]{Moore1}
{Moore} B., {Katz} N., {Lake} G., {Dressler} A., {Oemler} A., 1996, \nat, 379,
  613

\bibitem[{{Moore} {et~al}\mbox{.}(1998){Moore}, {Lake}, \& {Katz}}]{Moore2}
{Moore} B., {Lake} G., {Katz} N., 1998, \apj, 495, 139

\bibitem[{{Muzzin} {et~al}\mbox{.}(2013){Muzzin}, {Marchesini}, {Stefanon},
  {Franx}, {McCracken}, {Milvang-Jensen}, {Dunlop}, {Fynbo}, {Brammer},
  {Labbe}, \& {van Dokkum}}]{Muzzin}
{Muzzin} A. {et~al.}, 2013, ArXiv e-prints

\bibitem[{{Noeske} {et~al}\mbox{.}(2007){Noeske}, {Weiner}, {Faber},
  {Papovich}, {Koo}, {Somerville}, {Bundy}, {Conselice}, {Newman},
  {Schiminovich}, {Le Floc'h}, {Coil}, {Rieke}, {Lotz}, {Primack}, {Barmby},
  {Cooper}, {Davis}, {Ellis}, {Fazio}, {Guhathakurta}, {Huang}, {Kassin},
  {Martin}, {Phillips}, {Rich}, {Small}, {Willmer}, \& {Wilson}}]{Noeske}
{Noeske} K.~G. {et~al.}, 2007, \apjl, 660, L43

\bibitem[{{Oemler}(1974)}]{Oemler}
{Oemler}, Jr. A., 1974, \apj, 194, 1

\bibitem[{{Pasquali} {et~al}\mbox{.}(2010){Pasquali}, {Gallazzi}, {Fontanot},
  {van den Bosch}, {De Lucia}, {Mo}, \& {Yang}}]{Pasquali+10}
{Pasquali} A., {Gallazzi} A., {Fontanot} F., {van den Bosch} F.~C., {De Lucia}
  G., {Mo} H.~J., {Yang} X., 2010, \mnras, 407, 937

\bibitem[{{Peng} {et~al}\mbox{.}(2010){Peng}, {Lilly}, {Kova{\v c}},
  {Bolzonella}, {Pozzetti}, {Renzini}, {Zamorani}, {Ilbert}, {Knobel},
  {Iovino}, {Maier}, {Cucciati}, {Tasca}, {Carollo}, {Silverman}, {Kampczyk},
  {de Ravel}, {Sanders}, {Scoville}, {Contini}, {Mainieri}, {Scodeggio},
  {Kneib}, {Le F{\`e}vre}, {Bardelli}, {Bongiorno}, {Caputi}, {Coppa}, {de la
  Torre}, {Franzetti}, {Garilli}, {Lamareille}, {Le Borgne}, {Le Brun},
  {Mignoli}, {Perez Montero}, {Pello}, {Ricciardelli}, {Tanaka}, {Tresse},
  {Vergani}, {Welikala}, {Zucca}, {Oesch}, {Abbas}, {Barnes}, {Bordoloi},
  {Bottini}, {Cappi}, {Cassata}, {Cimatti}, {Fumana}, {Hasinger}, {Koekemoer},
  {Leauthaud}, {Maccagni}, {Marinoni}, {McCracken}, {Memeo}, {Meneux}, {Nair},
  {Porciani}, {Presotto}, \& {Scaramella}}]{Peng10}
{Peng} Y.-j. {et~al.}, 2010, \apj, 721, 193

\bibitem[{{Peng} {et~al}\mbox{.}(2012){Peng}, {Lilly}, {Renzini}, \&
  {Carollo}}]{Peng12}
{Peng} Y.-j., {Lilly} S.~J., {Renzini} A., {Carollo} M., 2012, \apj, 757, 4

\bibitem[{{Poggianti} {et~al}\mbox{.}(2013){Poggianti}, {Calvi}, {Bindoni},
  {D'Onofrio}, {Moretti}, {Valentinuzzi}, {Fasano}, {Fritz}, {De Lucia},
  {Vulcani}, {Bettoni}, {Gullieuszik}, \& {Omizzolo}}]{P+13}
{Poggianti} B.~M. {et~al.}, 2013, \apj, 762, 77

\bibitem[{{Poggianti} {et~al}\mbox{.}(2008){Poggianti}, {Desai}, {Finn},
  {Bamford}, {De Lucia}, {Varela}, {Arag{\'o}n-Salamanca}, {Halliday}, {Noll},
  {Saglia}, {Zaritsky}, {Best}, {Clowe}, {Milvang-Jensen}, {Jablonka},
  {Pell{\'o}}, {Rudnick}, {Simard}, {von der Linden}, \& {White}}]{P+08}
{Poggianti} B.~M. {et~al.}, 2008, \apj, 684, 888

\bibitem[{{Read} {et~al}\mbox{.}(2006){Read}, {Wilkinson}, {Evans}, {Gilmore},
  \& {Kleyna}}]{Read}
{Read} J.~I., {Wilkinson} M.~I., {Evans} N.~W., {Gilmore} G., {Kleyna} J.~T.,
  2006, \mnras, 366, 429

\bibitem[{{Salim} {et~al}\mbox{.}(2007){Salim}, {Rich}, {Charlot},
  {Brinchmann}, {Johnson}, {Schiminovich}, {Seibert}, {Mallery}, {Heckman},
  {Forster}, {Friedman}, {Martin}, {Morrissey}, {Neff}, {Small}, {Wyder},
  {Bianchi}, {Donas}, {Lee}, {Madore}, {Milliard}, {Szalay}, {Welsh}, \&
  {Yi}}]{Salim}
{Salim} S. {et~al.}, 2007, \apjs, 173, 267

\bibitem[{{Schiminovich} {et~al}\mbox{.}(2007){Schiminovich}, {Wyder},
  {Martin}, {Johnson}, {Salim}, {Seibert}, {Treyer}, {Budav{\'a}ri}, {Hoopes},
  {Zamojski}, {Barlow}, {Forster}, {Friedman}, {Morrissey}, {Neff}, {Small},
  {Bianchi}, {Donas}, {Heckman}, {Lee}, {Madore}, {Milliard}, {Rich}, {Szalay},
  {Welsh}, \& {Yi}}]{Schiminovich}
{Schiminovich} D. {et~al.}, 2007, \apjs, 173, 315

\bibitem[{{Scranton} {et~al}\mbox{.}(2002){Scranton}, {Johnston}, {Dodelson},
  {Frieman}, {Connolly}, {Eisenstein}, {Gunn}, {Hui}, {Jain}, {Kent},
  {Loveday}, {Narayanan}, {Nichol}, {O'Connell}, {Scoccimarro}, {Sheth},
  {Stebbins}, {Strauss}, {Szalay}, {Szapudi}, {Tegmark}, {Vogeley}, {Zehavi},
  {Annis}, {Bahcall}, {Brinkman}, {Csabai}, {Hindsley}, {Ivezic}, {Kim},
  {Knapp}, {Lamb}, {Lee}, {Lupton}, {McKay}, {Munn}, {Peoples}, {Pier},
  {Richards}, {Rockosi}, {Schlegel}, {Schneider}, {Stoughton}, {Tucker},
  {Yanny}, \& {York}}]{Scranton}
{Scranton} R. {et~al.}, 2002, \apj, 579, 48

\bibitem[{{Shin} {et~al}\mbox{.}(2008){Shin}, {Strauss}, {Oguri}, {Inada},
  {Falco}, {Broadhurst}, \& {Gunn}}]{Shin}
{Shin} M.-S., {Strauss} M.~A., {Oguri} M., {Inada} N., {Falco} E.~E.,
  {Broadhurst} T., {Gunn} J.~E., 2008, \aj, 136, 44

\bibitem[{{Simard} {et~al}\mbox{.}(2011){Simard}, {Mendel}, {Patton},
  {Ellison}, \& {McConnachie}}]{Simard}
{Simard} L., {Mendel} J.~T., {Patton} D.~R., {Ellison} S.~L., {McConnachie}
  A.~W., 2011, \apjs, 196, 11

\bibitem[{{Simard} {et~al}\mbox{.}(2002){Simard}, {Willmer}, {Vogt},
  {Sarajedini}, {Phillips}, {Weiner}, {Koo}, {Im}, {Illingworth}, \&
  {Faber}}]{GIM2D}
{Simard} L. {et~al.}, 2002, \apjs, 142, 1

\bibitem[{{Smith} {et~al}\mbox{.}(2009){Smith}, {Lucey}, \& {Hudson}}]{SLH}
{Smith} R.~J., {Lucey} J.~R., {Hudson} M.~J., 2009, \mnras, 400, 1690

\bibitem[{{Strateva} {et~al}\mbox{.}(2001){Strateva}, {Ivezi{\'c}}, {Knapp},
  {Narayanan}, {Strauss}, {Gunn}, {Lupton}, {Schlegel}, {Bahcall}, {Brinkmann},
  {Brunner}, {Budav{\'a}ri}, {Csabai}, {Castander}, {Doi}, {Fukugita}, {Gy{\H
  o}ry}, {Hamabe}, {Hennessy}, {Ichikawa}, {Kunszt}, {Lamb}, {McKay},
  {Okamura}, {Racusin}, {Sekiguchi}, {Schneider}, {Shimasaku}, \&
  {York}}]{Strateva}
{Strateva} I. {et~al.}, 2001, \aj, 122, 1861

\bibitem[{{Strauss} {et~al}\mbox{.}(2002){Strauss}, {Weinberg}, {Lupton},
  {Narayanan}, {Annis}, {Bernardi}, {Blanton}, {Burles}, {Connolly},
  {Dalcanton}, {Doi}, {Eisenstein}, {Frieman}, {Fukugita}, {Gunn},
  {Ivezi{\'c}}, {Kent}, {Kim}, {Knapp}, {Kron}, {Munn}, {Newberg}, {Nichol},
  {Okamura}, {Quinn}, {Richmond}, {Schlegel}, {Shimasaku}, {SubbaRao},
  {Szalay}, {Vanden Berk}, {Vogeley}, {Yanny}, {Yasuda}, {York}, \&
  {Zehavi}}]{Strauss}
{Strauss} M.~A. {et~al.}, 2002, \aj, 124, 1810

\bibitem[{{Taylor} {et~al}\mbox{.}(2010){Taylor}, {Franx}, {Brinchmann}, {van
  der Wel}, \& {van Dokkum}}]{Taylor}
{Taylor} E.~N., {Franx} M., {Brinchmann} J., {van der Wel} A., {van Dokkum}
  P.~G., 2010, \apj, 722, 1

\bibitem[{{Taylor} {et~al}\mbox{.}(2009){Taylor}, {Franx}, {van Dokkum},
  {Bell}, {Brammer}, {Rudnick}, {Wuyts}, {Gawiser}, {Lira}, {Urry}, \&
  {Rix}}]{Taylor2}
{Taylor} E.~N. {et~al.}, 2009, \apj, 694, 1171

\bibitem[{{Trujillo} {et~al}\mbox{.}(2006){Trujillo}, {F{\"o}rster Schreiber},
  {Rudnick}, {Barden}, {Franx}, {Rix}, {Caldwell}, {McIntosh}, {Toft},
  {H{\"a}ussler}, {Zirm}, {van Dokkum}, {Labb{\'e}}, {Moorwood},
  {R{\"o}ttgering}, {van der Wel}, {van der Werf}, \& {van
  Starkenburg}}]{Trujillo+06}
{Trujillo} I. {et~al.}, 2006, \apj, 650, 18

\bibitem[{{van den Bosch} {et~al}\mbox{.}(2008){van den Bosch}, {Aquino},
  {Yang}, {Mo}, {Pasquali}, {McIntosh}, {Weinmann}, \& {Kang}}]{vdB}
{van den Bosch} F.~C., {Aquino} D., {Yang} X., {Mo} H.~J., {Pasquali} A.,
  {McIntosh} D.~H., {Weinmann} S.~M., {Kang} X., 2008, \mnras, 387, 79

\bibitem[{{Wake} {et~al}\mbox{.}(2012{\natexlab{a}}){Wake}, {Franx}, \& {van
  Dokkum}}]{Wake2}
{Wake} D.~A., {Franx} M., {van Dokkum} P.~G., 2012{\natexlab{a}}, ArXiv
  e-prints

\bibitem[{{Wake} {et~al}\mbox{.}(2012{\natexlab{b}}){Wake}, {van Dokkum}, \&
  {Franx}}]{Wake1}
{Wake} D.~A., {van Dokkum} P.~G., {Franx} M., 2012{\natexlab{b}}, \apjl, 751,
  L44

\bibitem[{{Weinmann} {et~al}\mbox{.}(2009){Weinmann}, {Kauffmann}, {van den
  Bosch}, {Pasquali}, {McIntosh}, {Mo}, {Yang}, \& {Guo}}]{Weinmann}
{Weinmann} S.~M., {Kauffmann} G., {van den Bosch} F.~C., {Pasquali} A.,
  {McIntosh} D.~H., {Mo} H., {Yang} X., {Guo} Y., 2009, \mnras, 394, 1213

\bibitem[{{Wetzel} {et~al}\mbox{.}(2013){Wetzel}, {Tinker}, {Conroy}, \& {van
  den Bosch}}]{Wetzel}
{Wetzel} A.~R., {Tinker} J.~L., {Conroy} C., {van den Bosch} F.~C., 2013,
  \mnras, 432, 336

\bibitem[{{Willmer} {et~al}\mbox{.}(2006){Willmer}, {Faber}, {Koo}, {Weiner},
  {Newman}, {Coil}, {Connolly}, {Conroy}, {Cooper}, {Davis}, {Finkbeiner},
  {Gerke}, {Guhathakurta}, {Harker}, {Kaiser}, {Kassin}, {Konidaris}, {Lin},
  {Luppino}, {Madgwick}, {Noeske}, {Phillips}, \& {Yan}}]{Willmer}
{Willmer} C.~N.~A. {et~al.}, 2006, \apj, 647, 853

\bibitem[{{Wolf} {et~al}\mbox{.}(2009){Wolf}, {Arag{\'o}n-Salamanca}, {Balogh},
  {Barden}, {Bell}, {Gray}, {Peng}, {Bacon}, {Barazza}, {B{\"o}hm}, {Caldwell},
  {Gallazzi}, {H{\"a}u{\ss}ler}, {Heymans}, {Jahnke}, {Jogee}, {van Kampen},
  {Lane}, {McIntosh}, {Meisenheimer}, {Papovich}, {S{\'a}nchez}, {Taylor},
  {Wisotzki}, \& {Zheng}}]{Wolf}
{Wolf} C. {et~al.}, 2009, \mnras, 393, 1302

\bibitem[{{Woo} {et~al}\mbox{.}(2013){Woo}, {Dekel}, {Faber}, {Noeske}, {Koo},
  {Gerke}, {Cooper}, {Salim}, {Dutton}, {Newman}, {Weiner}, {Bundy}, {Willmer},
  {Davis}, \& {Yan}}]{Woo}
{Woo} J. {et~al.}, 2013, \mnras, 428, 3306

\bibitem[{{Yang} {et~al}\mbox{.}(2009){Yang}, {Mo}, \& {van den Bosch}}]{Yang3}
{Yang} X., {Mo} H.~J., {van den Bosch} F.~C., 2009, \apj, 695, 900

\bibitem[{{Yang} {et~al}\mbox{.}(2007){Yang}, {Mo}, {van den Bosch},
  {Pasquali}, {Li}, \& {Barden}}]{Yang1}
{Yang} X., {Mo} H.~J., {van den Bosch} F.~C., {Pasquali} A., {Li} C., {Barden}
  M., 2007, \apj, 671, 153

\bibitem[{{Yang} {et~al}\mbox{.}(2012){Yang}, {Mo}, {van den Bosch}, {Zhang},
  \& {Han}}]{YangDR7}
{Yang} X., {Mo} H.~J., {van den Bosch} F.~C., {Zhang} Y., {Han} J., 2012, \apj,
  752, 41

\bibitem[{{York} {et~al}\mbox{.}(2000){York}, {Adelman}, {Anderson},
  {Anderson}, {Annis}, {Bahcall}, {Bakken}, {Barkhouser}, {Bastian}, {Berman},
  {Boroski}, {Bracker}, {Briegel}, {Briggs}, {Brinkmann}, {Brunner}, {Burles},
  {Carey}, {Carr}, {Castander}, {Chen}, {Colestock}, {Connolly}, {Crocker},
  {Csabai}, {Czarapata}, {Davis}, {Doi}, {Dombeck}, {Eisenstein}, {Ellman},
  {Elms}, {Evans}, {Fan}, {Federwitz}, {Fiscelli}, {Friedman}, {Frieman},
  {Fukugita}, {Gillespie}, {Gunn}, {Gurbani}, {de Haas}, {Haldeman}, {Harris},
  {Hayes}, {Heckman}, {Hennessy}, {Hindsley}, {Holm}, {Holmgren}, {Huang},
  {Hull}, {Husby}, {Ichikawa}, {Ichikawa}, {Ivezi{\'c}}, {Kent}, {Kim},
  {Kinney}, {Klaene}, {Kleinman}, {Kleinman}, {Knapp}, {Korienek}, {Kron},
  {Kunszt}, {Lamb}, {Lee}, {Leger}, {Limmongkol}, {Lindenmeyer}, {Long},
  {Loomis}, {Loveday}, {Lucinio}, {Lupton}, {MacKinnon}, {Mannery}, {Mantsch},
  {Margon}, {McGehee}, {McKay}, {Meiksin}, {Merelli}, {Monet}, {Munn},
  {Narayanan}, {Nash}, {Neilsen}, {Neswold}, {Newberg}, {Nichol}, {Nicinski},
  {Nonino}, {Okada}, {Okamura}, {Ostriker}, {Owen}, {Pauls}, {Peoples},
  {Peterson}, {Petravick}, {Pier}, {Pope}, {Pordes}, {Prosapio},
  {Rechenmacher}, {Quinn}, {Richards}, {Richmond}, {Rivetta}, {Rockosi},
  {Ruthmansdorfer}, {Sandford}, {Schlegel}, {Schneider}, {Sekiguchi}, {Sergey},
  {Shimasaku}, {Siegmund}, {Smee}, {Smith}, {Snedden}, {Stone}, {Stoughton},
  {Strauss}, {Stubbs}, {SubbaRao}, {Szalay}, {Szapudi}, {Szokoly}, {Thakar},
  {Tremonti}, {Tucker}, {Uomoto}, {Vanden Berk}, {Vogeley}, {Waddell}, {Wang},
  {Watanabe}, {Weinberg}, {Yanny}, {Yasuda}, \& {SDSS Collaboration}}]{SDSS}
{York} D.~G. {et~al.}, 2000, \aj, 120, 1579

\end{thebibliography}

\appendix

\section{Appropriate Choice of Effective Radius} \label{scn:ressel}
\begin{figure*}
\centerline{	
  \includegraphics[clip=true,trim=0mm 0mm 0mm 0mm,scale=0.75,angle=0]{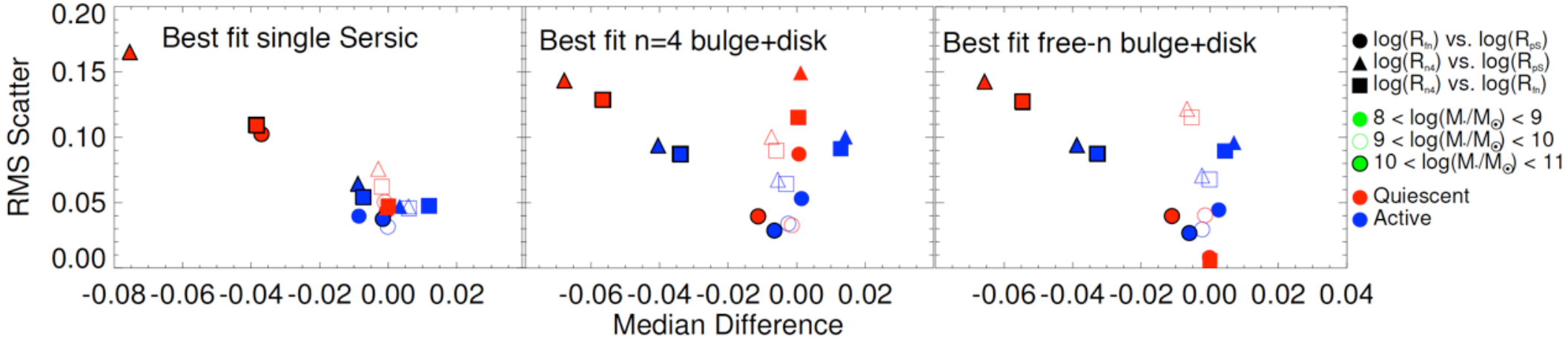}
		\label{fig:rleg}}
	\caption{Each point in these three plots represents a comparison between $\log R_e$ inferred from two model fits to a particular subpopulation of galaxies.  The median offset between the two measured is shown on the x-axis, and the {\it rms} scatter is given on the y-axis.  The legend at the right summarizes the different point styles.  Red and blue represent quiescent and active galaxies, respectively.  The symbol style (open, filled, or filled with a black outline) represents galaxies of increasing stellar mass (see green points in the legend for mass ranges).  The point shape indicates which two radii are being compared; the choices, increasing from simplest to most complex model, are $R_{pS}$ for a pure S\'{e}rsic fit, $R_{n4}$ for a bulge $+$ disk decomposition with the bulge index fixed to $n=4$, and $R_{fn}$ for a bulge $+$ disk decomposition model in which the bulge index $n$ is a free parameter.  The {\it left panel} shows galaxies for which a pure S\'{e}rsic model provides a good fit to the data; the {\it middle panel} shows those for which a $n=4$ bulge component is required, and the {\it right panel} is restricted to the 9 per cent of galaxies for which the free-$n$ bulge $+$ disk decomposition model is required.  
In all cases the radii compare well, relative to the sizes of the trends in $\log{R_e}$ we consider in this paper.  The largest offsets and variances are seen for the $n=4$ models, particularly at high masses.  
\label{fig:mrad}}
\end{figure*}
\citet{Simard} provide several different measures of galaxy radius, for each of the three types of model fits described in \S~\ref{scn:galprop}.  First, radii measured from the $r-$ and $g-$band images are provided; we use the $r-$band values as more representative of the stellar mass distribution.  Secondly, in addition to the circularized half-light ratio determined from curve-of-growth analysis, the catalogue provides the semi-major axis, computed from the model bulge and disk components as if their own semi-major axes were aligned.  The more physical option is not trivial to determine.  For disk-dominated galaxies, where the ellipticity is driven by inclination, the semi-major axis (independent of inclination for thin, circular disks) is probably the best choice.  However, for intrinsically elliptical galaxies the measured semi-major axis depends upon unknown orientation angle and is not always physically interesting.  We choose to use the circularized half-light radius because of its simplicity and the fact that it is commonly used amongst other observers.  We show below that our results are not particularly sensitive to this choice (see Figure~\ref{fig:rsmafn}).

The most critical decision to be made was which of the three model fits to use.
\citet{Simard} use an F-test to determine which model best fits the data.  From these, they provide two parameters, P$_{pS}$ and P$_{n4}$, which respectively give the probability that a bulge $+$ disk decomposition is not needed (compared to a pure S\'{e}rsic model) and the probability that a free n bulge $+$ disk decomposition is not needed (compared to an n = 4 bulge $+$ disk model).  We take their recommendation of setting a threshold value for P$_{pS}$ and P$_{n4}$ of 0.32 (1$\sigma$) for distinguishing which model fits best.  Using this criteria, 74$\%$ of galaxies are best fit by a pure S\'{e}rsic model (P$_{pS}$ $>$ 0.32), 17$\%$ of galaxies are best fit by a de Vaucouleurs (n $=$ 4) bulge $+$ disk (P$_{pS}$ $\leq$ 0.32 and P$_{n4}$ $>$ 0.32), and only 9$\%$ were best fit by free n bulge $+$ disk (P$_{pS}$ $\leq$ 0.32 and P$_{n4}$ $\leq$ 0.32).

We divided our galaxy sample into three subsets depending on which model fits best, and examined how the circularized radii depend on the choice of model for each subset, as a function of the stellar mass of the galaxy.  We further consider active and quiescent galaxies separately, since they have fundamentally different relations between mass and size.  To quantify this comparison, for each galaxy type, and for 1 dex wide bins in $\log(M_*/M_\Sun)$, we compare the
radius measured from one fit against each of the other two, and consider trends in bias and scatter.

The results are summarized in  Figure~\ref{fig:mrad}, where 
each point indicates the median difference between two radius measurements (on the x-axis) and the {\it rms} scatter in that difference (y-axis).  Different symbols distinguish between active/quiescent galaxies (blue/red colours), different stellar masses (open, filled, and outlined-filled symbols ranging from low to high mass), and which radii are being compared (triangles, squares and circles as described in the legend).  In each panel, the comparisons are restricted to the subsample in which one of the three models is the statistically preferred fit, as indicated in the plot titles.  

The first general point to make is that in all cases the average difference between two measurements of $\log({R_e})$ is $<0.08$, and the {\it rms} scatter is $<0.12$; this is small relative to all the trends we consider and we can expect none of our results will be sensitive to the choice of radius, as we show explicitly below.  Moreover, the fits from the simplest, pure S\'{e}rsic model and those with the most complex, free-$n$ bulge $+$ disk decomposition model, give very similar results for $R_e$, with a median difference of $\lesssim 0.2$ and an {\it rms} scatter of $\lesssim 0.06$.   This is true regardless of which fit is statistically preferred.  The biggest differences are observed between the radius inferred from the $n=4$ bulge $+$ disk decomposition model (R$_{n4}$) and either of the other two (R$_{pS}$, triangles or R$_{fn}$, squares).  This is most pronounced for the more massive galaxies (outlined, filled points) and is again not very dependent on which model fit is actually preferred.

\begin{figure*}
\centerline{
	\subfigure{
		\includegraphics[clip=true,trim=0mm 0mm 0mm 0mm,scale=1,angle=0]{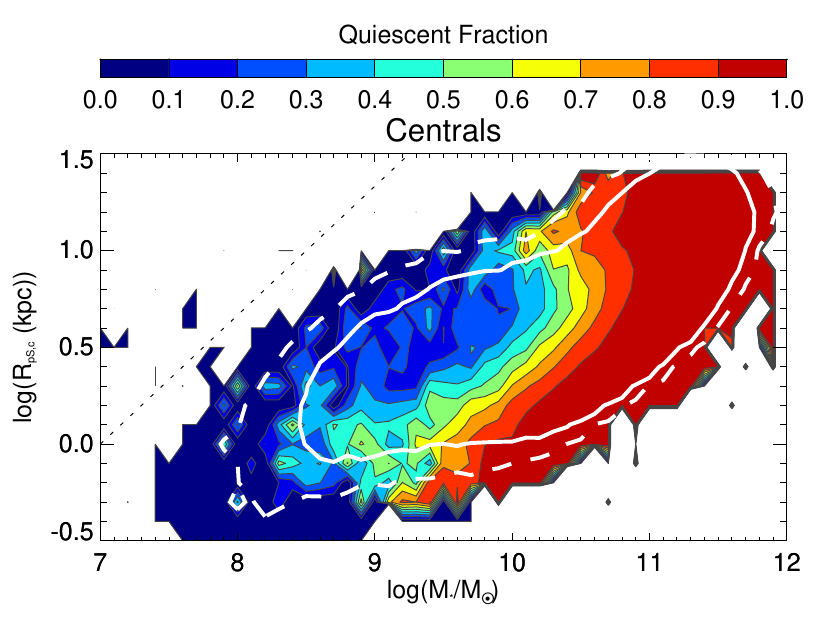}
		\label{fig:cadat}}
	\subfigure{
		\includegraphics[clip=true,trim=0mm 0mm 0mm 0mm,scale=1,angle=0]{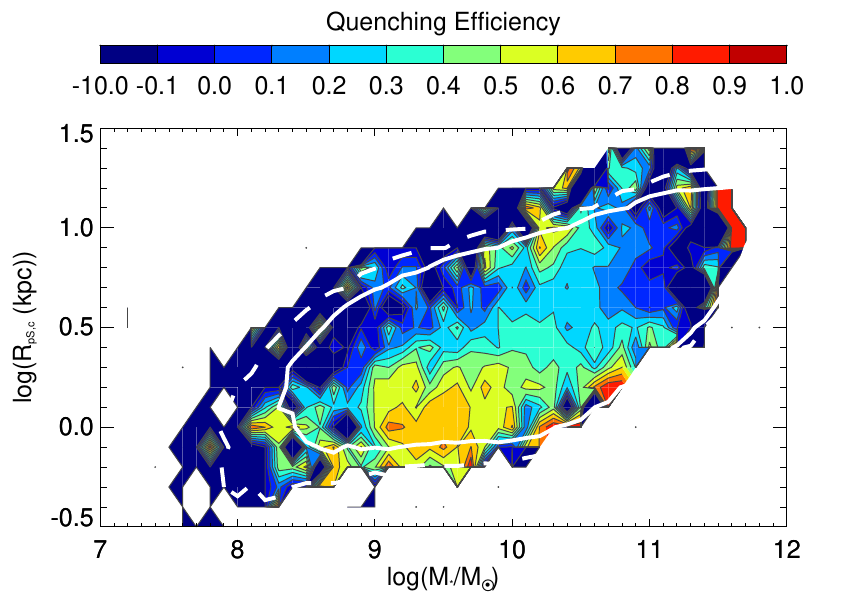}
		\label{fig:sadat}}}
	\caption{The quiescent fraction of central galaxies (left), and the quenching efficiency (right), are shown as a function of stellar mass M$_*$ and pure S\'{e}rsic model circularized radius R$_{pS,c}$. These compare well with our fiducial results, presented in Figures~\ref{fig:csqf} (\textit{left}) and \ref{fig:qeff}.  }
\label{fig:rcps}
\end{figure*}

\begin{figure*}
\centerline{
	\subfigure{
		\includegraphics[clip=true,trim=0mm 0mm 0mm 0mm,scale=1,angle=0]{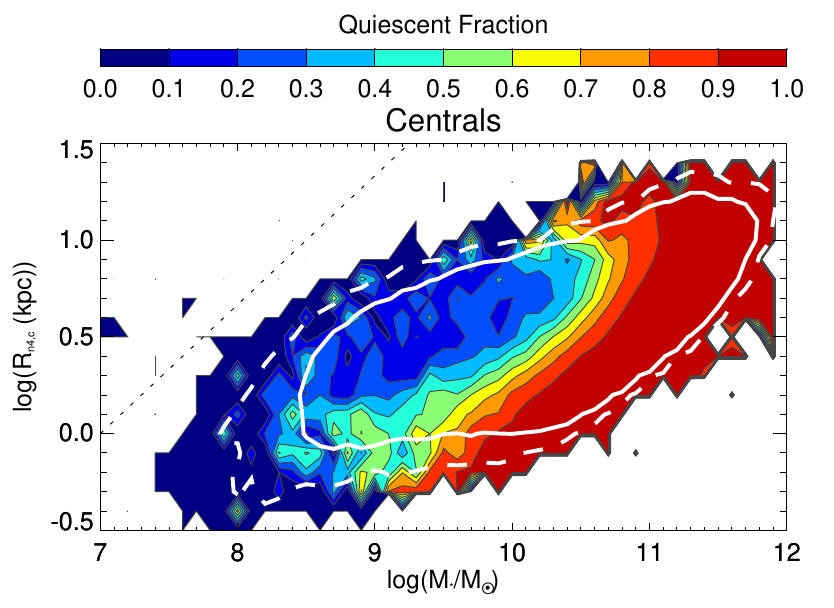}
		\label{fig:cadat2}}
	\subfigure{
		\includegraphics[clip=true,trim=0mm 0mm 0mm 0mm,scale=1,angle=0]{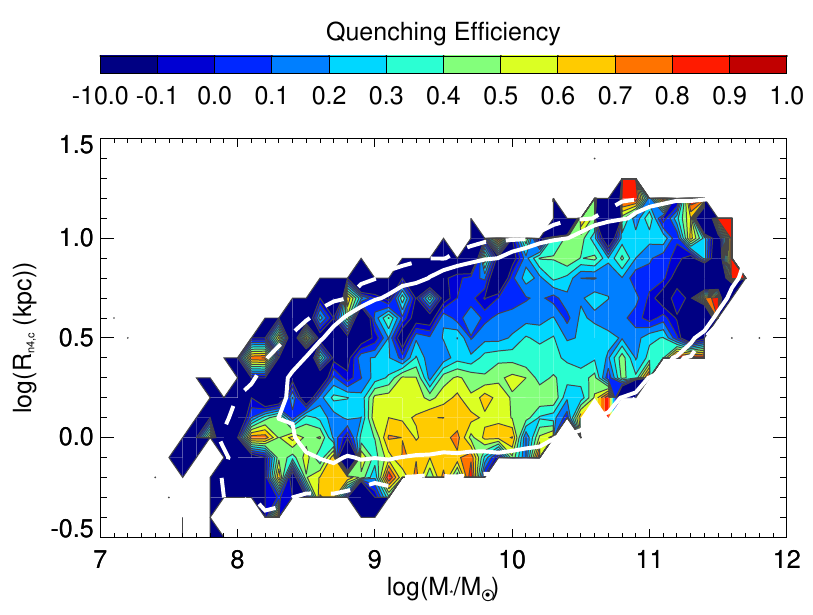}
		\label{fig:sadat2}}}
	\caption{As Figure~\ref{fig:rcps}, but where the y-axis is the circularized radius from the  n$_b=$4 bulge $+$ disk decomposition, R$_{n4,c}$.}
\label{fig:rcn4}
\end{figure*}

\begin{figure*}
\centerline{
	\subfigure{
		\includegraphics[clip=true,trim=0mm 0mm 0mm 0mm,scale=1,angle=0]{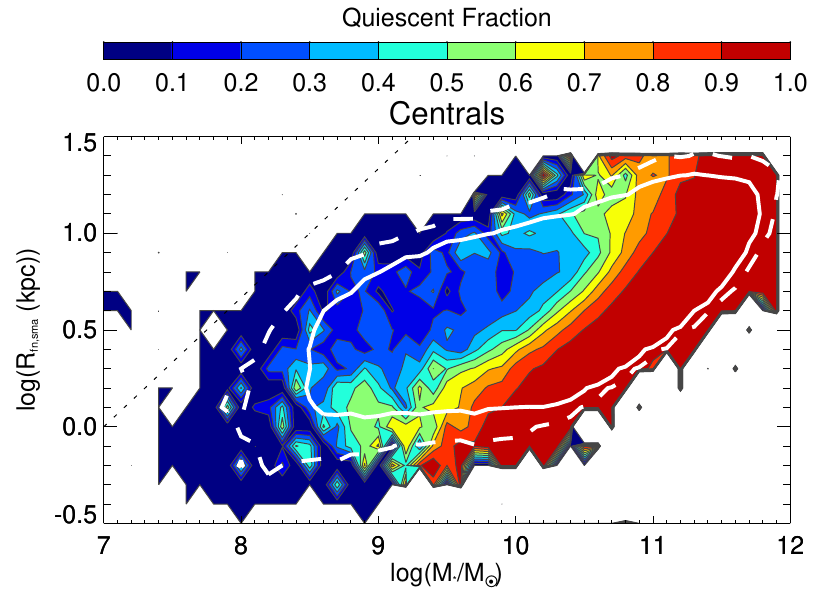}
		\label{fig:cadat3}}
	\subfigure{
		\includegraphics[clip=true,trim=0mm 0mm 0mm 0mm,scale=1,angle=0]{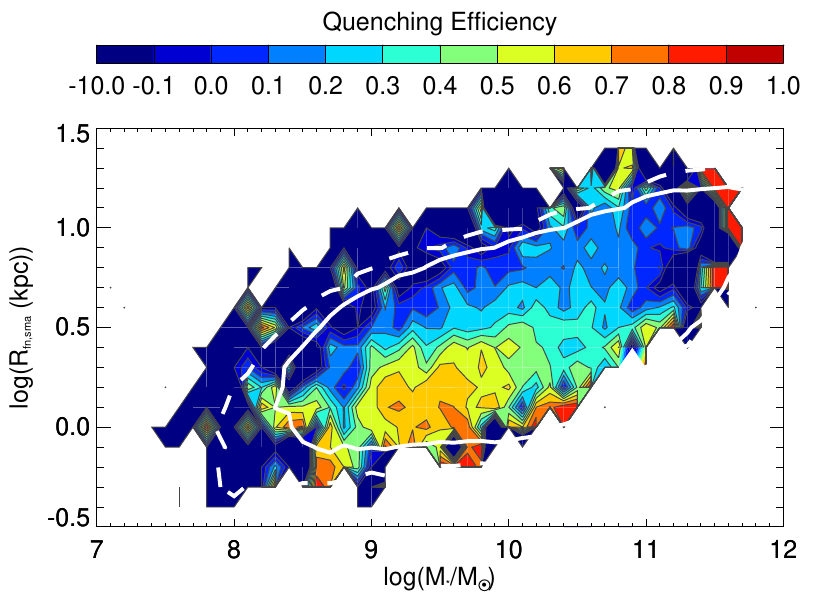}
		\label{fig:sadat3}}}
	\caption{As Figure~\ref{fig:rcps}, but where the y-axis is the semi-major axis from our fiducial model (free$-n$ bulge $+$ disk decomposition model), R$_{fn,sma}$.}
\label{fig:rsmafn}
\end{figure*}

In Figure~\ref{fig:rcps} we show the quiescent fraction of central galaxies, and the quenching efficiency, where the effective radius is measured from the pure, S\'{e}rsic model circularized radii.  This can be compared with our results presented in Figures~\ref{fig:csqf} (\textit{left}) and \ref{fig:qeff}.  Similarly, Figure~\ref{fig:rcn4} shows the result when the effective radius is measured from the $n$ $=$ $4$ bulge + disk model fit.  It is evident from these figures that none of our conclusions are dependent on the choice of $R_e$.  The main difference is apparent in the behaviour of the largest ($R_e>10$kpc), most massive ($M_*>10^{10.5} M_\Sun$) central galaxies, and these are rare (c.f. Figure~\ref{fig:wdc}).  

Finally, in Figure~\ref{fig:rsmafn} we show the same results computed for our fiducial model choice (free-$n$ bulge + disk decomposition), but where we use the semi-major axis rather than the circularized half-light radius.  Again the results are almost unchanged, apart from the most massive, largest galaxies. 

In summary, the measurement of half-light radius is not strongly dependent on the model choice.  The main results of this paper are that the contours of quiescent fraction follow $M_*$ $\propto$ $ R_e^{1.5}$ over most of the parameter space, and that the quenching efficiency is not uniform, but peaks at masses and sizes where satellites are most common.  These results are completely unchanged when different choices of $R_e$ are made. 

\section{Stellar Mass, Inferred Velocity Dispersion and Surface Density Distributions} \label{scn:IVDSDF}
\begin{figure*}
\centerline{     
       \subfigure{
               \includegraphics[clip=true,trim=0mm 0mm 0mm 0mm,scale=1.,angle=0]{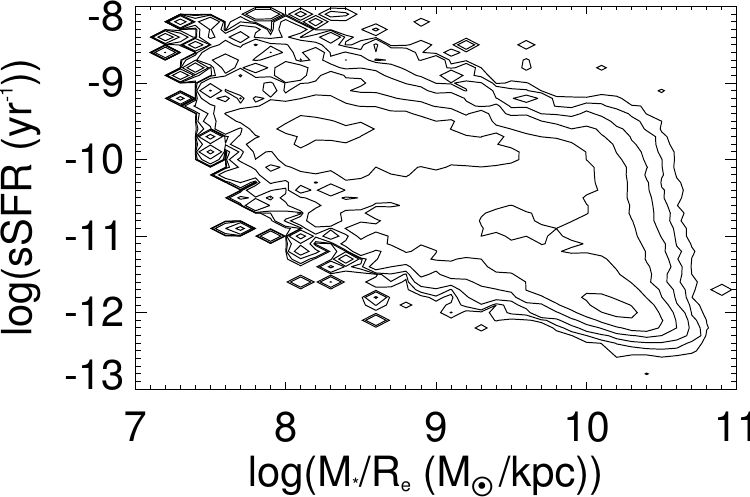}
               \label{fig:ivdidt}}
       \subfigure{
               \includegraphics[clip=true,trim=0mm 0mm 0mm 0mm,scale=1.,angle=0]{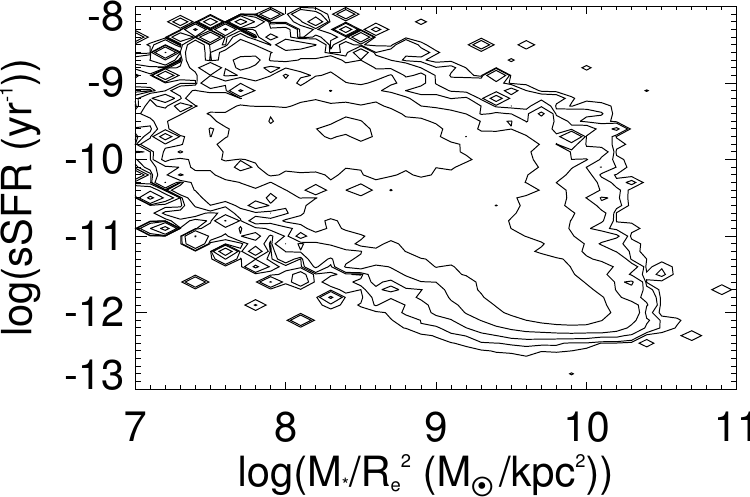}
                \label{fig:sdidt}}}
        \caption{The correlation between sSFR and  inferred velocity dispersion (left), and surface density (right) are shown as contours representing the weighted data distribution.  The bimodality of the population is apparent in both cases.}
\label{fig-ssfr-app}
\end{figure*}

We start by showing the distribution of sSFR as a function of inferred velocity dispersion M$_*$/R$_e$  and surface density M$_*$/R$_e^2$, in Figure~\ref{fig-ssfr-app}.  The bimodality seen in Figure~\ref{fig:idf} is readily apparent in these presentations, as well.  Since our division between quiescent and active galaxies is weakly dependent on stellar mass, it does not correspond to a line in these figures.  

\begin{figure*}
\centerline{
        \subfigure{
                \includegraphics[clip=true,trim=0mm 0mm 0mm 0mm,scale=0.75,angle=0]{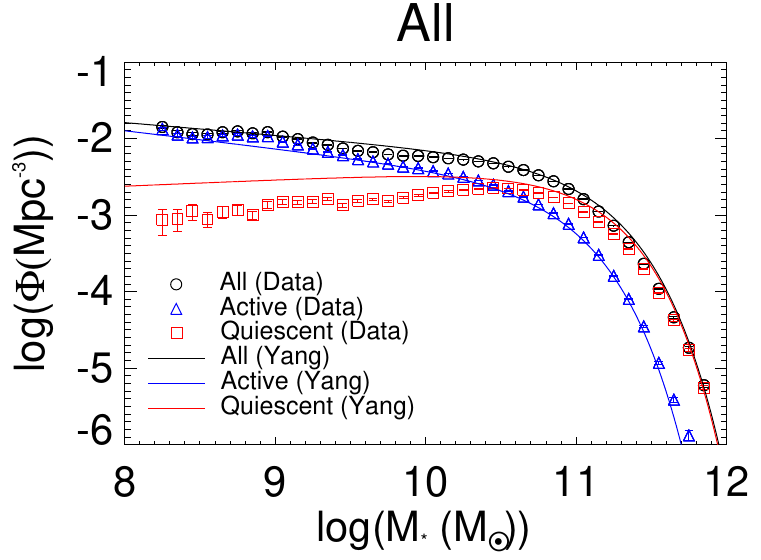}
                \label{fig:SMFall}}
        \subfigure{
              \includegraphics[clip=true,trim=0mm 0mm 0mm 0mm,scale=0.75,angle=0]{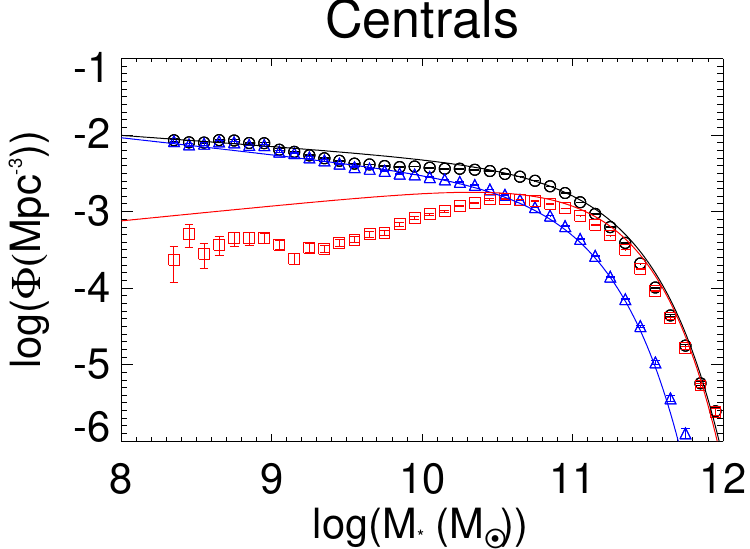}
                \label{fig:SMFc}}
        \subfigure{
                \includegraphics[clip=true,trim=0mm 0mm 0mm 0mm,scale=0.75,angle=0]{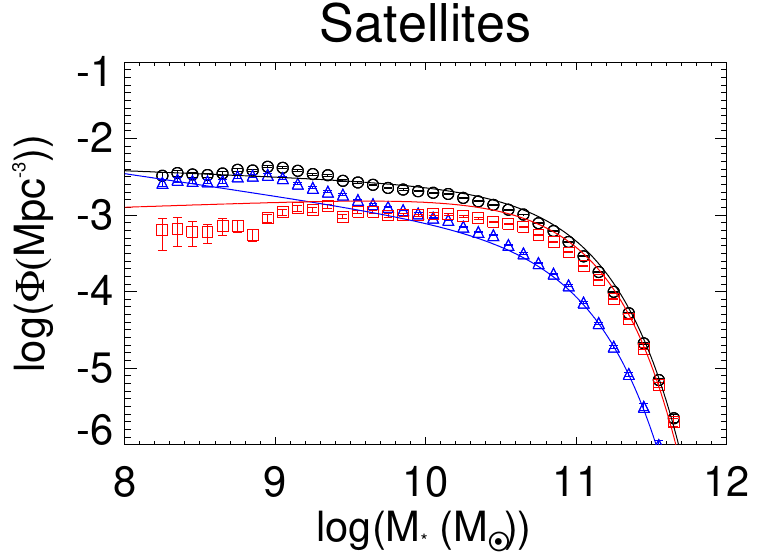}
                \label{fig:SMFs}}}
        \caption{Stellar mass functions are presented for all galaxies (left), centrals (centre), and satellites (right).  In each panel we show the mass function for active galaxies (blue), and quiescent galaxies (red), and for the full population in black.  Schechter function fits from \citet{Yang3} are also shown, for comparison. Since their classification of quiescent and active galaxies was based on a single colour, rather than SFR, our mass functions for these subpopulations differ from theirs.}
\label{fig:smf}
\end{figure*}

We use the definition of central and satellite galaxy from \citet{Yang3}, but our choice of active and quiescent is different from theirs.  Thus, to enable comparison we show the stellar mass functions for the subpopulations considered in this paper, in Figure~\ref{fig:smf}.  These are compared with the  Schechter function fits provided by 
\citet{Yang3}.  We closely recover the total mass function of centrals and satellites reported by \citet{Yang3}.  However, by basing our definition of quiescence on sSFR rather than colour only, we reduce the number of low-mass, quiescent galaxies.

\begin{figure*}
\centerline{
        \subfigure{
                \includegraphics[clip=true,trim=0mm 0mm 0mm 0mm,scale=0.8,angle=0]{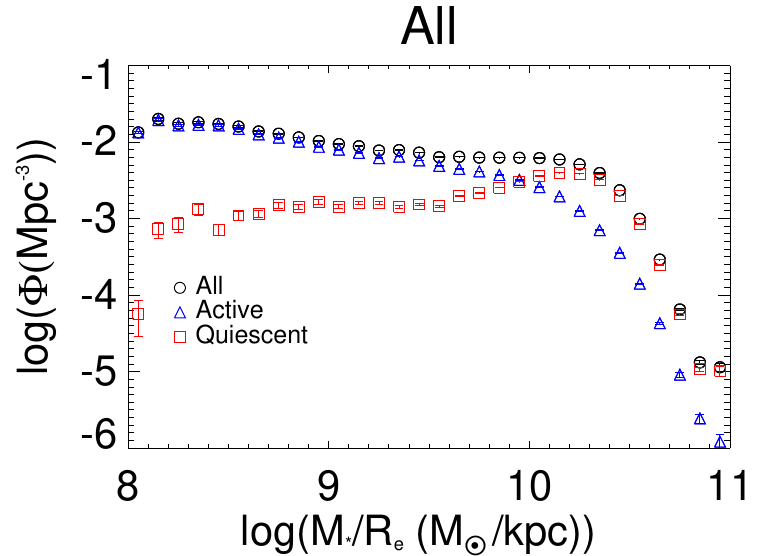}
                \label{fig:IVDFall}}
        \subfigure{
                \includegraphics[clip=true,trim=0mm 0mm 0mm 0mm,scale=0.8,angle=0]{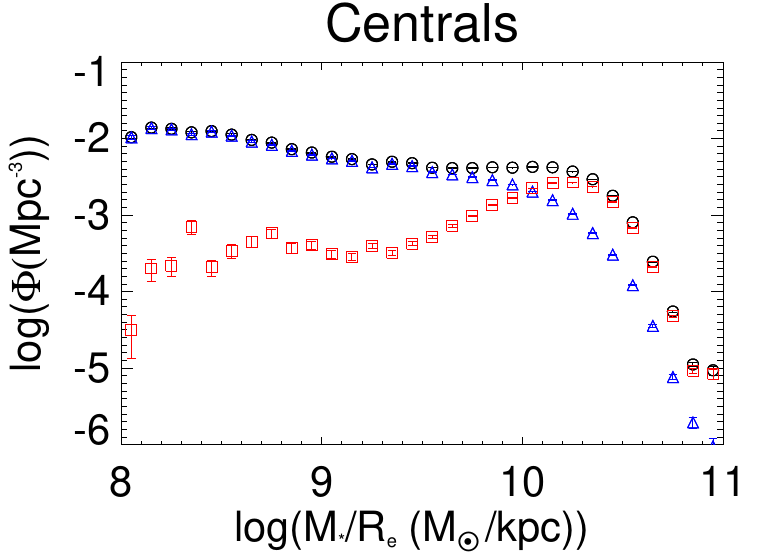}
                \label{fig:IVDFc}}
        \subfigure{
                \includegraphics[clip=true,trim=0mm 0mm 0mm 0mm,scale=0.8,angle=0]{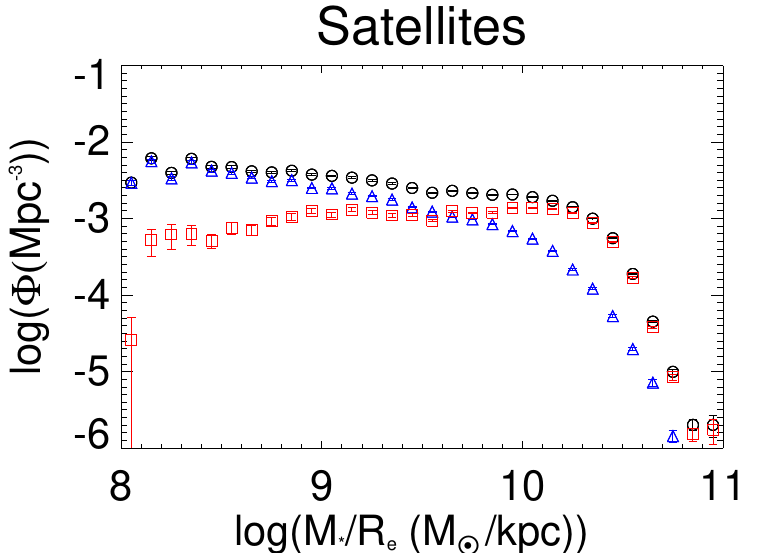}
                \label{fig:IVDFs}}}
        \caption{Inferred Velocity Dispersion functions for all galaxies (left), centrals (centre), and satellites (right), split into all galaxies (black), active galaxies (blue), and quiescent galaxies (red).  }
\label{fig:ivdf}
\end{figure*}

\begin{figure*}
\centerline{
        \subfigure{
                \includegraphics[clip=true,trim=0mm 0mm 0mm 0mm,scale=0.8,angle=0]{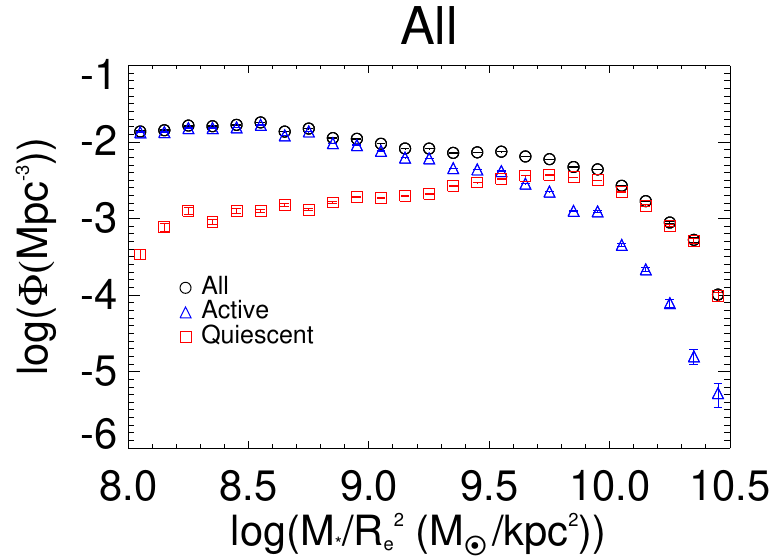}
                \label{fig:SDFall}}
        \subfigure{
                \includegraphics[clip=true,trim=0mm 0mm 0mm 0mm,scale=0.8,angle=0]{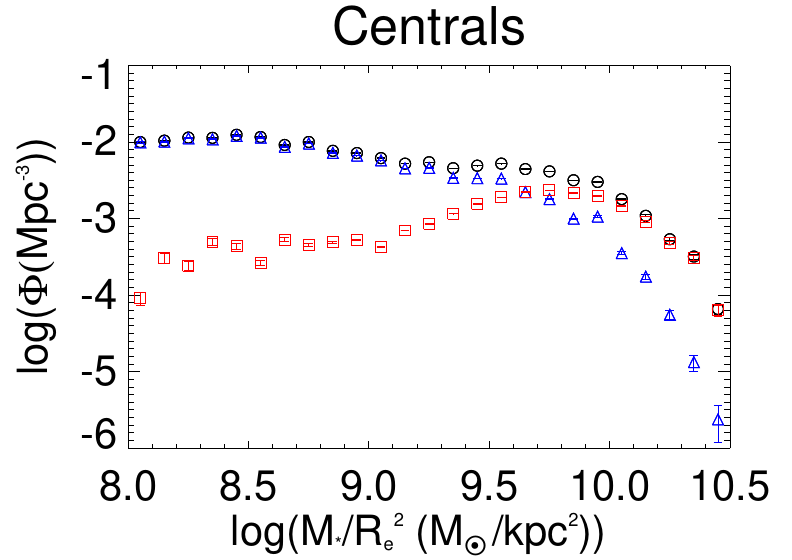}
                \label{fig:SDFc}}
        \subfigure{
                \includegraphics[clip=true,trim=0mm 0mm 0mm 0mm,scale=0.8,angle=0]{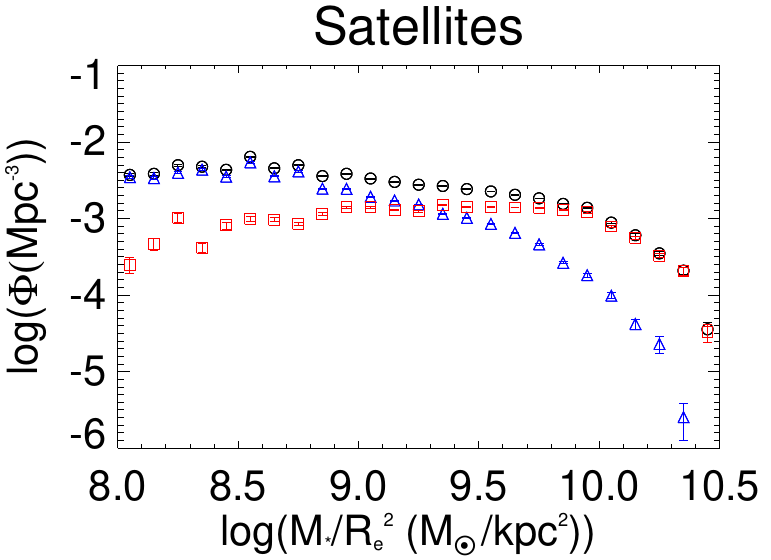}
                \label{fig:SDFs}}}
        \caption{Surface Density functions for all galaxies (left), centrals (centre), and satellites (right), split into all galaxies (black), active galaxies (blue), and quiescent galaxies (red).}
\label{fig:sdf}
\end{figure*}

It is also instructive to see the number density functions for the inferred velocity dispersion (Figure~\ref{fig:ivdf}) and surface density (Figure~\ref{fig:sdf}).  In each case we separate the population into centrals and satellites, and into active and quiescent galaxies.  Note that the completeness limits of these samples are not trivial to determine, as they are partly limited by the resolution limit of SDSS. Figure~\ref{fig:wdc} shows that the abundance of active galaxies is maximal near the limits of the survey ($M_*=10^8M_\Sun$ and $R_e=0.3$ kpc) and thus the sample may not be purely mass-limited.  As lines of constant $M_*/R_e^2$ lie nearly parallel to the star--forming galaxy relation, this means we may only be strictly mass-limited for the highest- and lowest-surface brightnesses.

\end{document}